\documentclass{aastex631}
\usepackage{threeparttable}

\newcommand{\eg}{{\it e.g.}\rm}
\newcommand{\ie}{{\it i.e.}\rm}

\submitjournal{Planetary Science Journal}
\accepted{January 20, 2022}

\begin{document}
\title{Large Impacts onto the Early Earth: Planetary Sterilization and Iron Delivery}

\author[0000-0001-8920-0356]{Robert I. Citron}
\affiliation{Department of Earth and Planetary Science, University of California Davis, Davis, California, USA}

\author[0000-0001-9606-1593]{Sarah T. Stewart}
\affiliation{Department of Earth and Planetary Science, University of California Davis, Davis, California, USA}


%
%


\begin{abstract}
Late accretion onto the Hadean Earth included large impacts that could have influenced early habitability, either by sterilizing the planet or alternatively catalyzing the origin of life by delivering iron required to create a reducing environment/atmosphere. We present 3D numerical simulations of 1500-3400 km diameter impacts on the early Earth in order to quantify their effects on planetary habitability. We find sterilizing impact events require larger projectiles than previously assumed, with a 2000-2700 km diameter impactor required to completely melt Earth's surface and an extrapolated $>$700 km diameter impactor required for ocean-vaporization. We also find that reducing environments are less likely to arise following large impacts than previously suggested, because $>$70\% of the projectile iron is deposited in the crust and upper mantle where it is not immediately available to reduce surface water and contribute to forming a reducing atmosphere. Although the largest expected late accretion impacts ($\sim$1 lunar mass) delivered sufficient iron to the atmosphere to have reduced an entire ocean mass of water, such impacts would also have melted the entire surface, potentially sequestering condensing iron that is not oxidized quickly. The hypothesis that life emerged in the aftermath of large impacts requires an efficient mechanism of harnessing the reducing power of iron sequestered in the crust/mantle, or an origin of life pathway that operates in more weakly-reducing post-impact environments that require smaller quantities of impact-delivered iron.

\end{abstract}

\section{Introduction}

\begin{figure}[h!]
\centering
\includegraphics[width=0.9\textwidth,trim={1.0cm 7cm 4cm 1.5cm},clip]{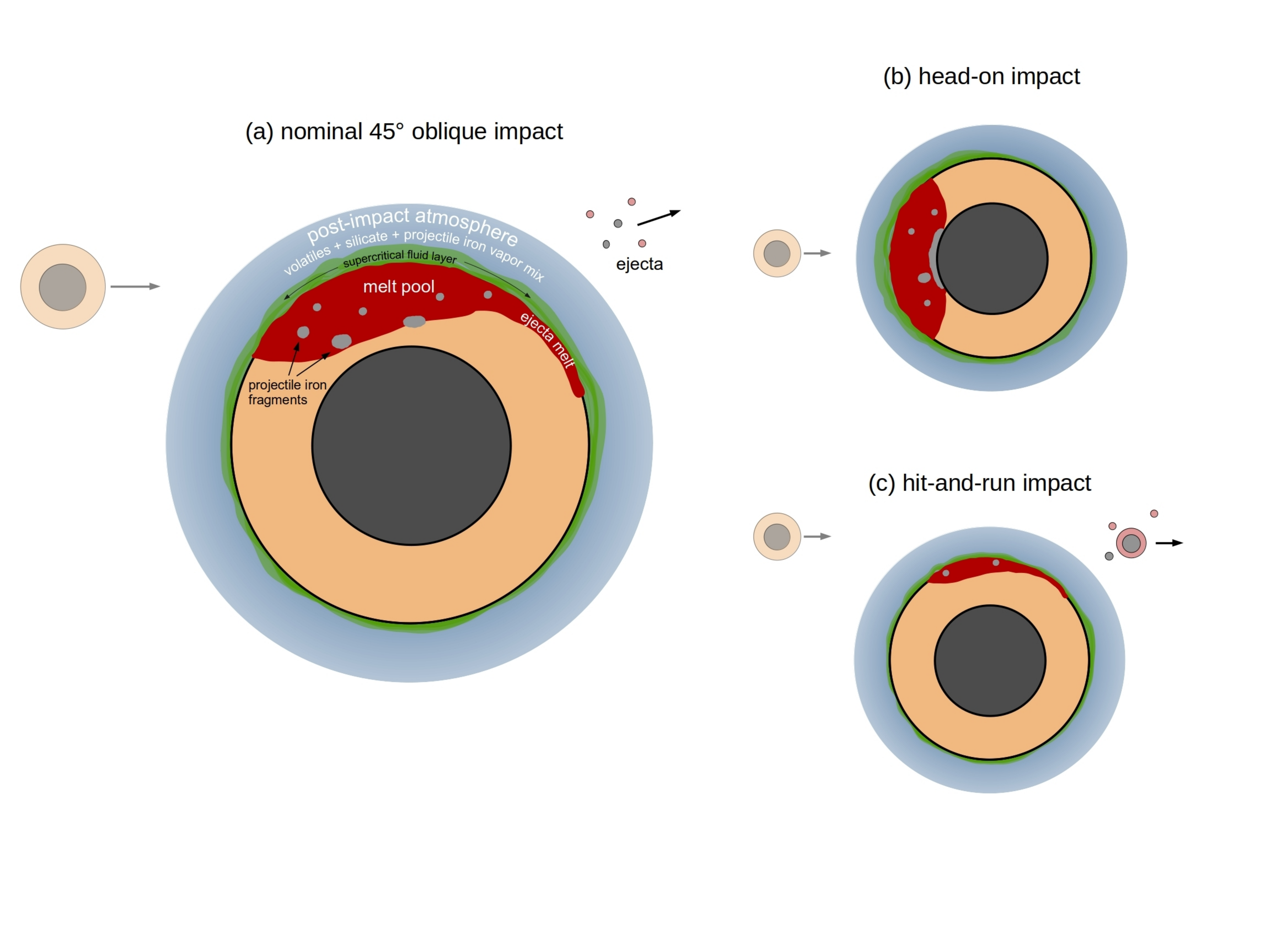}
\caption{Habitability of Earth in the aftermath of late accretion impacts depends strongly on the post-impact melt and iron distribution, and the energy deposited into the post-impact atmosphere. (a) In a nominal oblique impact a portion of the mantle is melted and ejecta heating also occurs downrange of the impact. The post-impact Earth has an ambiguous surface boundary of supercritical fluid, and the atmosphere consists of volatiles and a mix of vaporized silicate and projectile iron. Projectile iron is deposited in the melt/mantle/atmosphere, and some fraction also escapes the system with the impact ejecta. (b) In a head-on impact the projectile iron penetrates deep into the mantle, potentially reaching the core for larger impacts, and little material escapes the system. (c) In hit-and-run impacts, less melt is generated by the impact and a significant fraction of the projectile escapes the system. }
\label{fig:diagram}
\end{figure}

During the tail-end of accretion, the Hadean Earth likely experienced a large number of impacts that could have greatly inhibited or increased early habitability. Sufficiently large impacts could have sterilized the planet, for example by globally melting the outer crust or vaporizing an early ocean \citep[\eg,][]{Sleep1989,Benner2020}, with the timing of the last sterilizing impact placing a limit on the origin of the ancestors to present-day life. 
However, it is also possible that early planetary impacts helped generate an environment favorable to the origin of life. 
Although the early Earth likely had a weakly reducing atmosphere, rich in CO$_2$, H$_2$O, and N$_2$ \citep{poole1951evolution,Holland1962,abelson1966chemical,Kasting2014}, planetary impacts could have generated transient strongly reducing atmospheres by delivering metallic iron that would have reduced pre-existing surface water/oceans \citep[\eg,][]{Genda2017,Benner2020,Zahnle2020}. Reducing atmospheres, rich in H$_2$ and CH$_4$, can rapidly generate ribonucleic acid (RNA) precursor compounds, making iron delivery from impacts a potentially critical process to the origin of life. 
Early habitability may therefore depend on the planetary effects of large impacts during the Hadean (Figure \ref{fig:diagram}, a period where late accretion bombardment likely consisted of multiple projectiles 200$-$1000 km in diameter and potentially included impactors as large as 3000 km in diameter \citep[\eg,][]{Bottke2010,Morbidelli2018}.

Despite the importance of large impacts to early habitability, their efficiency in melting the surface and delivering iron has only been explored by a limited number of numerical modeling efforts \citep[\eg,][]{Svetsov2005,Genda2017,Marchi2017}. Because of the sparse exploration of this impact parameter space with numerical models, studies estimating the effect of planetary impacts on sterilization and habitability have needed to make generalized assumptions for how energy and iron are distributed during impacts. Previous works have assumed that 25$-$50\% of the projectile kinetic energy is partitioned towards evaporating a pre-existing ocean (via radiative heat from the post-impact rock-vapor atmosphere), and that all of the projectile iron is available to reduce pre-existing surface water \citep[\eg,][]{Sleep1989,Zahnle2020}. 
These rough estimates are only partially supported numerical models \citep[\eg,][]{Svetsov2005,Genda2017,Marchi2017}, and limitations in the scope of prior impact studies makes it difficult to assess the comprehensive effects of large-scale 500$-$3000 km diameter impacts on habitability. 
For example, while \citet{Svetsov2005} conducted a set of simulations in the relevant size regime, the models were in 2D and used undifferentiated projectiles and therefore could not be used to test the effects of impact angle or assess the post-impact distribution of iron. \citet{Genda2017} and \citet{Marchi2017} provide a more complete examination of the disruption of iron projectile cores in 3D simulations, but do not fully examine the post-impact spatial distribution of melt or the creation of rock vapor atmospheres. A wider range of impact simulations is therefore required to comprehensively examine the effects of impacts on habitability in terms of iron delivery, ocean vaporization, and surface melting. 

In this work, we conduct a set of 3D simulations of large impacts on the early Earth in order to better understand what types of impacts would either sterilize the early Earth or provide sufficient iron to sustain a post-impact reducing atmosphere. 
We examine impacts 1500$-$3400 km in diameter over a range of impact velocities and angles in order to fully understand the effects of these impact parameters on surface melting, ocean vaporization, and the post-impact distribution of iron. We also utilize updated equations of state for the iron core and forsterite mantle that better estimate the behavior of these materials during large-scale impacts \citep{Stewart2019,stewart2020zenodo-ironalloy}. In particular, these updated equations of state provide improved estimates of material vaporization, which is essential for modeling the formation of post-impact atmospheres. 
We find that sterilizing impacts, although larger than previously estimated, can be sufficiently small to have occurred multiple times during late accretion. For example, Ceres-scale projectiles could have vaporized several ocean masses of water (although ocean-vaporization may not be completely sterilizing), and most half-lunar mass projectiles could have caused global surface melting. We also find that the delivery of projectile iron to locations where it can react with surface water is inefficient; only the largest impacts (3400 km diameter) deliver large amounts of iron to the surface and atmosphere, and much of this iron could be entrained in post-impact surface melt prior to reacting with surface water. Most projectile iron is delivered to the mantle, where it is unclear to what extent (and over what timescales) it would react with surface water or carbon in the mantle. While sterilizing impacts may be common during late accretion, forming transient reduced atmospheres from impact-delivered iron may require an efficient mechanism of harnessing the reducing power of iron sequestered in the upper crust and mantle.

\section{Background} \label{sec:background}

\subsection{Late Accretion}
The amount of material delivered to the Earth during late accretion (referring to material added after the Moon-forming impact) is constrained by the observed mantle abundances of highly siderophile elements (HSEs). 
Because HSEs should have partitioned into the core during differentiation, the excess of HSEs in Earth's mantle suggests that Earth accreted $\sim$0.5$-$1 wt\% of material with a bulk chondritic composition after core segregation, often referred to as the \textit{late veneer} \citep[\eg,][]{Chou1978,Day2007,Bottke2010,Morbidelli2015,Brasser2016}. 
However, the late veneer mass inferred from the mantle HSE excess (equivalent in scale to a single chondritic impactor of diameter $D\sim 2300-2900$ km) is only a lower limit for the total amount of material delivered during late accretion; inefficient delivery of HSEs to the mantle during late accretion impacts (\eg, due to large fractions of projectile metals being ejected or merging with the Earth's core) could result in a late accretion mass 2$-$5 times greater than normally assumed based on the observed mantle HSE enrichment \citep{Marchi2017}. 
Assuming late accretion was similar in mass to the late veneer constraint and had a size distribution similar to the inner asteroid belt, most of the mass would have been concentrated in the largest impactor ($D>2500$ km) \citep{Bottke2010}. 
Concentration of the late accretion mass in a small number of large impactors might explain the enhancement of HSE abundances in the Earth's mantle relative to the lunar mantle, because a small number of large impactors would have preferentially hit the Earth due to its larger gravitational cross-section \citep{Bottke2010}. 

It should be noted, however, that there is an alternative model that suggests the HSE excess in the Earth's mantle came mostly from the core of the Moon-forming impactor \textit{Theia} \citep{Newsom1989,Sleep2016}. While simulations of the Moon-forming impact suggest that the post-impact terrestrial mantle would have been mostly molten \citep{Canup2001}, allowing HSEs to be effectively sequestered in the Earth's core, only a few percent of \textit{Theia's} core is required to be retained in the Earth's mantle to explain the terrestrial HSE excess. Incorporation of a small fraction of \textit{Theia's} core into the terrestrial mantle is feasible if the material was directly mixed into the post-impact terrestrial mantle (via rapid oxidation of iron and retention of the corresponding HSEs in the mantle) or was ejected into the proto-lunar disk before falling back to Earth \citep{Sleep2016}. This model is not generally emphasized because chondritic HSE proportions are observed not only in Earth's mantle but in the Moon, Mars, Vesta, and the Angrite parent body, so there appears to be a late added chondritic component to all differentiated bodies \citep{Dale2012}. Additionally, the low HSE abundances in the lunar mantle relative to the Earth might also be explained by inefficient retention of material during lunar impacts \citep{Kraus2015,Zhu2019}.

There are generally two scenarios for the context of late accretion for the Earth-Moon system: a \textit{cataclysm} scenario in which there is a surge in the impact rate around 3.9 Ga, or an \textit{accretion tail} scenario in which bombardment declines monotonically with time \citep[][and references therein]{Morbidelli2018}. 
In the cataclysm scenario, a late dynamical instability in the giant planets scatters trans-Neptunian objects, asteroids, and leftover planetesimals into the inner solar system, resulting in a spike in the impact rate that can explain the apparent concentration of impact basins on the Moon with ages $\sim$ 3.9 Ga, commonly referred to as the lunar late heavy bombardment \citep[\eg,][]{Gomes2005}.
Alternatively in the accretion tail scenario, bombardment of the terrestrial planets is dominated by planetesimals leftover from planet formation, which are gradually removed via collisions and dynamical effects; a cataclysm resulting from a giant planet instability is still possible, but it would have occurred early in Solar System evolution, meaning observations indicating a lunar late heavy bombardment are a result of sampling biases or age resetting \citep[\eg][]{hartmann1975lunar,boehnke2016illusory,zellner2017cataclysm,Miljkovic2013}.
In the context of habitability, both scenarios are similar because the total mass delivered is constrained by the excess HSE budget of the terrestrial mantle \citep{Morbidelli2018}, and differ mainly in the timing of bombardment. In both cases, bombardment of the Earth by such a large amount of material (including projectiles $>$2500 km diameter) would have significantly affected early habitability, either by sterilizing the surface or generating reducing atmospheres via the delivery of iron.

\subsection{Planetary Sterilization}

The timing of the last sterilizing impact sets a limit on the earliest age for the origin of the precursors to present-day life \citep{Sleep1989,Benner2020,Zahnle2020}. Due to the general decrease in impactor size over time, the last sterilizing impact is best constrained by estimating the ``minimum sterilizing impact'', the smallest scale impact that would make Earth uninhabitable. A lower limit for the minimum sterilizing impact is given by the scale of impact required to vaporize the early ocean \citep{Sleep1989}. Although such an impact may not wipe out every single organism, if the early ecosystem was reliant on a surface ocean then vaporization of all surface water could constitute a sterilizing event because inhospitable conditions at the surface would persist for thousands of years post-impact \citep{Sleep1989}. A more conservative sterilization estimate requires vaporization of only the upper 200 m of the ocean (the photic zone), which would destroy an early ecosystem reliant on photosynthetic biota \citep{Sleep1989}. However, depending on the early ecosystem, ocean vaporization may not constitute a sterilizing event because subsurface life could survive. Additionally, because the seismic wave generated by an impact rapidly traverses the globe, environments may be buried and thermally protected from short-lived atmospheric heating, allowing lifeforms to persist in habitable pockets even if the post-impact surface is otherwise uninhabitable. Complete sterilization can only be ensured by globally melting the Earth's surface and upper crust, providing an upper limit for the minimum sterilizing impact. A more conservative upper limit would be an impact that heats the upper crust above the limit for hyperthermophiles (80-110$^\circ$C) \citep{Abramov2009}. While the true threshold for planetary sterilization is unclear, if no impacts that globally melted the surface occurred during late accretion, it is possible that Earth could have been continuously habitable since the Moon-forming impact and experienced only partially-sterilizing events.

The minimum scale impact that could vaporize an ocean is unclear. While an impact would immediately vaporize surface water in the impacted region, global ocean vaporization is most likely caused by the envelopment of the Earth in an impact-generated hot silicate vapor atmosphere \citep{Sleep1989,Zahnle2020}. The post-impact rock vapor atmosphere would radiate both outwards into space and downwards towards the surface at a temperature at least 1500 K. If the silicate vapor atmosphere has sufficient total internal energy, its thermal radiation towards the surface can completely vaporize any pre-existing ocean. \citet{Sleep1989} and \citet{Zahnle2020} suggest that impacts of diameter 350-440 km could vaporize an entire ocean mass of water (one present day surface ocean mass = $1.4\times10^{21}$kg), based on the assumption that 25-50\% of the impact kinetic energy is partitioned towards vaporizing surface water; however, their assumed energy partitioning is a rough estimate that requires validation with comprehensive numerical impact simulations.

One of the few studies to numerically examine the effects of impacts on ocean vaporization was conducted by \citet{Svetsov2005}, who modeled a suite of 2D impacts 500 to 3000 km in diameter at an impact velocity of 15 km/s. \citet{Svetsov2005} found that only 4$-$10 \% of the projectile's kinetic energy was partitioned into the internal energy of the rock vapor atmosphere, less than half the 25$-$50\% value used by \citet{Sleep1989} and \citet{Zahnle2020}. However, a value closer to 25\% can be obtained when considering other sources of vaporization such as ejecta deposition; \citet{Svetsov2005} found that 12$-$27\% of the projectile kinetic energy was deposited in the ejecta layer, suggesting that a 500 km diameter impact is sufficient to vaporize an entire ocean mass of water, only slightly larger than the 440 km impactor suggested by \citet{Sleep1989}. However, because \cite{Svetsov2005} only modeled impacts in 2D and reported global averages, the spatial distribution of the deposited ejecta energy in more realistic oblique collisions is unclear. This is particularly important for smaller impacts that result in proximal ejecta deposition instead of a thick global ejecta layer. \citet{Svetsov2005} also modeled undifferentiated impactors, which underestimate the impact kinetic energy for a given projectile diameter.

The minimum impact necessary to globally melt the Earth's surfaces is also poorly constrained. Traditional estimates of impact melt are derived from fitting power-law scaling relations to numerical simulations (half space models) of small-scale ($\sim$ 10 km diameter) impacts \citep[\eg,][]{Pierazzo1997a,Pierazzo2000a,Barr2011}. However, these estimates do not match impact melt volumes from simulations of larger scale impacts ($>$ 100 or 400 km diameter) \citep{Marchi2014,Monteux2016}. Additionally, most impact generated melt is confined to the impact basin, and it is unclear how far the impact melt would spread from extrusion during post-impact isostatic rebound \citep[\eg,][]{Tonks1993,ivanov2003impacts}. \citet{Marchi2014} estimate that for impacts $>$ 100 km in diameter, the region covered in melt $>$ 3 km thick is about 20-30 times the impact diameter, requiring a 1300-2000 km diameter impact to completely melt the Earth's surface. This is generally consistent with the results of \citet{Svetsov2005}, which found that a 2000 km diameter impact was required to cause significant (kilometers deep) global melting. Global surface melting can also be caused by the deposition of impact ejecta. For example, \citet{Svetsov2005} estimate that a 500 km diameter impact would result in a global layer of molten ejecta with an average thickness of 200 m. However, without 3D simulations of the post-impact ejecta distribution it is unclear if such an impact would constitute a global sterilization event or if there would be local regions of unmelted crust.

A more conservative estimate of sterilization from impact heating is given by the scale of impact necessary to heat the upper crust above the habitable limit for hyperthermophiles (80-110$^\circ$C) \citep{Abramov2009}. \citet{Abramov2009} examined the spatial distribution of subsurface regions where temperatures remained within habitable limits during a stochastic half-space model of the planetary bombardment, finding that large portions of the subsurface could have remained habitable during a late heavy bombardment scenario. However, \citet{Abramov2009} focused on habitability during a late cataclysm (late heavy bombardment) where the impactor population is based on a spike in the impactor flux $\sim$ 3.9 Ga inferred from the lunar basin record and is less massive than the late accretion mass constrained by mantle HSE enrichment. \citet{Abramov2009} therefore use a maximum impactor diameter of 300 km, significantly smaller than the $>$1000 km diameter impactors expected to constitute late accretion as constrained by HSE enrichment. Models of a bombardment more representative of late accretion suggest that almost all of Earth's original crust would have been reprocessed by impact-induced mixing, melt, and burial during late accretion \citep{Marchi2014}. In practice, applying the sterilization criteria of \citet{Abramov2009} to 3D simulations of larger scale impacts is difficult because global impact models are often computationally restricted to surface grid sizes of 50-100 km, significantly larger in scale than potential habitable zones in the upper 4 km of crust.

\subsection{Reducing post-impact atmospheres}

In contrast to potentially sterilizing the early Earth, large impacts have also been proposed as a critical catalyst for the origin of life \citep[\eg][]{Benner2020}. In particular, impacts may have been critical to generating an environment suitable for the abiological formation of RNA, which would have paved the way for an``RNA-First" origin of life in which the earliest lifeforms used self-replicating RNA molecules for genetics in an``RNA World". The development of an RNA World requires the formation of a large number of RNA precursor compounds, which are generally reduced compounds that form in atmospheres with abundant reducing gases. The formation of a reducing atmosphere in the Hadean is difficult because Earth's mantle is expected to to have been oxidized to a state near the fayalite-magnetite-quartz buffer (FMQ) relatively quickly after core formation. However, planetary impacts during late accretion could deliver metallic iron capable of forming a transient reducing atmosphere in the Hadean favorable for the formation of RNA precursor compounds \citep{Benner2020,Zahnle2020}. 

Reduction of surface water with iron proceeds via the reaction Fe $+$ H$_2$O $\rightarrow$ FeO $+$ H$_2$. 
For a late accretion of 0.5 wt\% Earth's mass, if the majority of the material was delivered in a single impactor with core mass fraction of 0.3 it would correspond to $\sim10^{22}$ kg of iron, sufficient to reduce 2.3 ocean masses of water \citep{Zahnle2020}. The production of such a large amount of H$_2$ (several tens of bars), which would react to form other reducing gases such as CH$_4$, should facilitate the formation of sufficient RNA precursor compounds for an RNA World \citep{Zahnle2020,Benner2020}. However, it is unclear if the majority of the impactor iron would be available to react with the pre-existing surface water, which would be in a steam or supercritical fluid state post-impact, because much of the iron delivered in large impacts may be sequestered into the mantle.

\citet{Genda2017} and \citet{Marchi2017} both conducted detailed examination of the distribution of iron following large, late-accretion type impacts. \citet{Genda2017} studied the effects of a 3000 km diameter projectile impacting the early Earth, specifically examining the mechanical separation of the core during the impact. \citet{Genda2017} found that for a 45$^\circ$ impact, 60\% of the projectile core would be available to react with water based on the amount of bound iron that experienced a sufficiently high strain rate to fragment during the impact. However, it is unclear in \citet{Genda2017} how much of the fragmented iron is delivered to the surface or mixed reactively with water during the impact as opposed to being sequestered in the mantle. Similar simulations by \citet{Marchi2017} show that a significant fraction of the projectile core ($\sim$ 60\%) would be entrained in the Earth's mantle. Because prior studies of the distribution of iron delivered in late accretion impacts have not fully distinguished between iron delivered to the surface and post-impact atmosphere and iron delivered to the mantle, further work is required to estimate the amount of reduced gases that could be produced from the interaction of impact delivered iron and surface water.

\section{Methods}\label{sec:methods}

To assess the effects of impacts on early habitability, we conducted a suite of 3D numerical simulations of late accretion impacts. Impacts were modeled using the GADGET-2 smooth particle hydrodynamics (SPH) code \citep{Springel05}, which was modified to utilize tabulated equations of state to model planetary collisions \citep{marcus2009collisional,marcus2011role}. 
The differentiated Earth and projectiles were modeled using the Analytic Equations of State (ANEOS) code package \citep{Thompson1972,Melosh2007,Thompson2019,Stewart2019report} using the modifications included in M-ANEOS v1 \citep{Thompson2019}. Mantle material was modeled using an updated version of forsterite that provides a better fit to experimental data in much of the regime of large impacts \citep{stewart2019zenodo-forsterite,Stewart2019}. Iron cores were modeled as Fe$_{85}$Si$_{15}$ iron alloy and constrained by recent experimental results \citep{stewart2020zenodo-ironalloy}. The equation of state tables were bilinearly interpolated. We use an artificial viscosity parameters $\alpha$=0.8, within the typical values of 0.5-1.0, and $\beta=2\alpha$ \citep{Springel05}
In this initial work, we did not include the effects of material strength.

To encompass the range of expected impacts, we modeled projectiles of mass 0.0012, 0.003, 0.006, and 0.012 $\textrm{M}_{\textrm{Earth}}$ ($D$ = 1500, 2000, 2700, or 3400 km), corresponding to tenth, quarter, half, and full lunar mass objects. 
The impact velocity $v_{imp}$ was set to 1.1, 1.5, or 2 $v_{esc}$, where $v_{esc}$ is the two-body escape velocity of the projectile and the Earth, as constrained by the expected impact velocity of Earth-crossing asteroids \citep{LeFeuvre2008}. 
The impact angle $\theta$ was set to 0, 30, 45, or 60$^\circ$, where 0$^\circ$ corresponds to a head-on impact. The head-on cases were run in order to compare with prior work conducted in 2D \citep[\eg][]{Svetsov2005}. The impactors and Earth were initialized using WoMa (World Maker) \citep{Ruiz-Bonilla2021}, a wrapper for the SEAGen stretched equal-area particle generator \citep{Kegerreis2019}. Particles within each body have slightly variable masses so that all particles have a density within 1\% of the desired value \citep{Kegerreis2019}, which for our targets and projectiles results in a maximum variation of 8\% between particles within each body. Constructing the target and projectile from discrete shells also results in differences in the average particle mass between the target and projectile particles; in our simulations the target and projectile particles differ in mass by a maximum of 30\%. The Earth was initialized with an isentropic thermal profile for the solid material with a specific entropy of 1800 J kg$^{-1}$ K$^{-1}$ for iron and 2810 J kg$^{-1}$ K$^{-1}$ for forsterite, corresponding to a mantle potential temperature of $\sim$ 1900 K, and projectiles were initialized with an adiabatic temperature profile corresponding to the same surface temperature. The projectiles and Earth were given core mass fractions of 0.33 and 0.32, respectively. Prior to the full impact simulations, all bodies were run in isolation for up to 12 hours of relaxation time, so that the rms velocity of the SPH particles converged to a small initial value. 
Due to computational constraints, simulations were limited to between 5$\times10^{5}$ and 3$\times10^{6}$ total SPH particles, so that there were at least 4000 particles in the projectile regardless of its size. 
All impact models were run for 24 hours of simulation time. 

\begin{figure}[h!]
\centering
\includegraphics[width=0.9\textwidth,trim={0 0cm 0 0cm},clip]{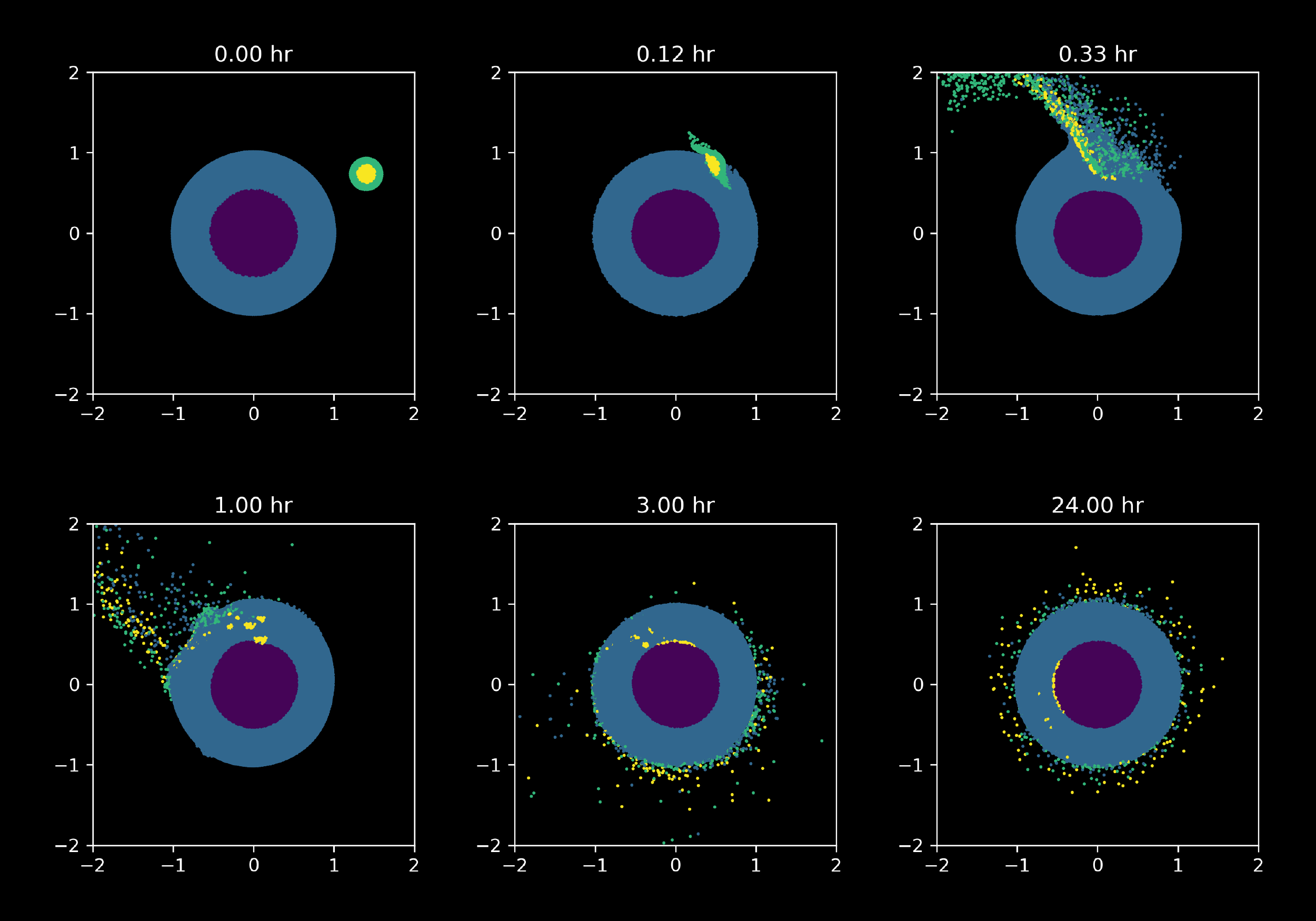}
\caption{Snapshots of a simulation of a 0.006 M$_{Earth}$ projectile impacting a Earth target at 1.5 $v_{esc}$ and $\theta=45^{\circ}$. Colors track the projectile and target core and mantle materials. Axis units are in present-day Earth radii (6371 km). }
\label{fig:example_simulation}
\end{figure}

Snapshots from an example simulation are shown in Figure \ref{fig:example_simulation}, which highlights how our oblique impact simulations resulted in downrange ejecta deposition and the production of a silicate vapor atmosphere that encircled the Earth by the end of the simulation. In our strengthless simulations, the core of the projectile quickly sank through the warm, fluid earth and merged with the Earth's core. However, this is likely a result of our models neglecting material strength. Simulations with material strength are expected to result in a significant portion of the projectile core becoming embedded in the Earth mantle because the mantle is not melted all the way to the Earth's core, and will be a focus of future work. Because of this, we do not make a distinction between projectile core material delivered to the Earth's core versus the mantle, instead treating both the core and mantle as the `interior'. We note, however, that even if iron was initially embedded at the base of the impact melt pool and above the solid mantle, the large deposits of iron would likely sink to the core either as descending diapirs or via the propagation of iron filled dikes \citep{Tonks1992,Rubie2015}. 

In our analysis, we classify SPH particles at the end of each simulation as either `interior', `surface', `atmosphere', 'disk' or `escaped'. We define the planetary surface as particles with P $>$ 10 GPa and $\rho$ $<$ 1000 kg/m$^3$, except for super-critical fluid particles which are partitioned between the surface and atmosphere using a more advanced scheme. This generally encompasses the outer one or two layers of SPH particles, effectively defining the outer layer of the final planet. However, defining the interface between the surface and atmosphere is difficult because in SPH simulations there is a smooth transition in density instead of a well-defined surface-atmosphere particle density contrast; additionally, in high energy impacts a near-surface layer of super-critical fluid is formed that could be categorized as either surface or atmosphere. We classify a near-surface super-critical fluid particle as a surface particle if its effective physical radius (particle mass divided by particle density) overlaps with any other surface particle's effective physical radius; otherwise, the super-critical fluid particle is classified as an atmosphere particle. 
Interior particles are all particles located below the surface, and escaped particles are particles with velocities greater than the two body escape velocity. Particles above the surface that are bound are divided into the atmosphere and disk. The boundary between the atmosphere and disk is also ambiguous because there is a smooth transition in particle properties. We estimate the inner radius of the disk as the location where there is a peak in the kinetic energy (in the x-y plane) of particles \citep{Stewart2021}.
According to these definitions, the post-impact atmosphere consists primarily of both pure vapor particles and supercritical fluid particles, but also include particles in a mixed vapor+liquid or vapor+solid state. 

The phase of each SPH particle is determined using the ANEOS phase boundaries in Pressure-Entropy space. The mass fraction of liquid, vapor, or solid for mixed-phase states is determined using the lever rule with specific entropy as the independent variable. A particle is supercritical fluid if its pressure and temperature are greater than the respective critical point values. 
The GADGET-2 simulations conserved energy during each model run, with the total change in internal, gravitational, and kinetic energy at the end of the simulations differing from the initial kinetic energy by an average of $\sim$1\%. Figure \ref{fig:energyhist} shows the exchange between kinetic, internal, and gravitational potential energy, and the conservation of total energy, for an example simulation.

\section{Results}

\subsection{Sterilization}\label{sec:results-sterilization}

\begin{figure}[h!]
\centering
\includegraphics[width=0.9\textwidth,trim={0 0cm 0 0.5cm},clip]{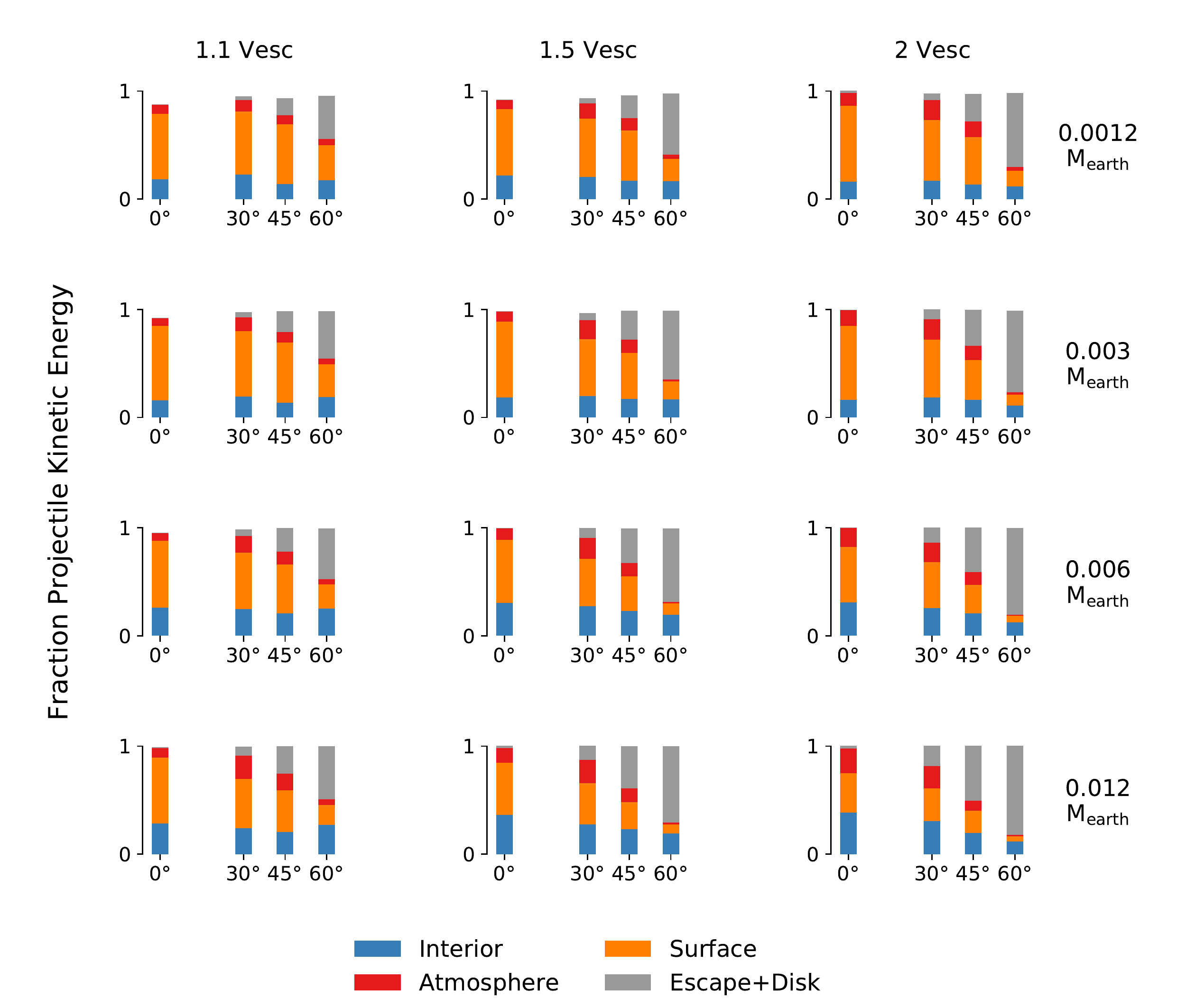}
\caption{Stacked bar charts showing how the initial projectile kinetic energy is distributed between the interior, surface, atmosphere, and escaping plus disk material, after 24 hrs of simulation time. Charts are distributed in columns and rows according to the projectile velocity and mass. The x-axis in each bar chart shows simulations with different impact angles (0$^\circ$ = head-on collision). See Table \ref{tbl:energy_part}.}
\label{fig:energy_partitioning}
\end{figure}

For each simulation, we quantify the potential of the impact to sterilize the early Earth either through vaporization of a pre-existing ocean or globally melting the surface. Both of these processes depend on how the projectile kinetic energy is partitioned during the impact, in particular the amount of energy partitioned into heating the Earth's surface or deposited into the impact-generated atmosphere. The post-impact energy partitioning for our simulations is shown in Figure \ref{fig:energy_partitioning}, with the interior, surface, atmosphere, and escape+disk reservoirs defined as in Section \ref{sec:methods}. For most simulations, the majority of the projectile's kinetic energy is deposited into the surface and interior. However, for more oblique or higher velocity impacts, the majority of the projectile's kinetic energy escapes the system as unbound ejecta due to the `hit-and-run' nature of these collisions. 

\newpage
\subsubsection{Ocean vaporization}

We assume that the onset of global ocean vaporization results primarily from thermal radiation of the impact-generated atmosphere downward onto the surface \citep{Sleep1989}. While this may not be the primary cause of ocean vaporization in large impacts that globally melt the surface, it is likely the dominant process in smaller impacts where the impact-heated surface melt is localized to a small portion of the planet's surface area. Therefore, in order to estimate the minimum scale impact required to vaporize a preexisting ocean, we examine if the hot rock vapor atmosphere that envelopes the planet post-impact has sufficient thermal energy to globally vaporize surface water.

We computed the energy in the atmosphere available to vaporize ocean water as the increase in internal energy of particles classified as atmospheric particles at the end of each simulation (i.e., the sum of each atmospheric particle's final internal energy minus its initial internal energy). This approximates the excess heat added to the atmospheric particles that can be radiated away as the atmosphere cools. We find that 1-21\% of the projectile kinetic energy is partitioned into the internal energy of the post-impact atmosphere (Table \ref{tbl:energy_part}). This is somewhat less than the 25-50\% estimate used by \citet{Zahnle2020} and \citet{Sleep1989} for the fraction of projectile kinetic energy partitioned towards ocean vaporization. However, it should be noted that our estimate does not include heating of the surface due to ejecta deposition or the hot iron component of the atmosphere, which would also have contributed to ocean vaporization. Given the internal energy of the post-impact atmosphere (Table \ref{tbl:energy_part}), simple radiative cooling with an effective temperature of 2300 K suggests the silicate atmosphere should cool within $\sim$ 1 to 100 yrs. A more complete radiative transfer model by \citet{Svetsov2005} suggests the silicate vapor atmosphere should last $\sim$ 10 yrs for a 3000 km diameter projectile impacting head-on at 15 km/s.

Given that the energy required to vaporize one ocean mass of water is $5\times10^{27}$ J, all of our simulations deposited sufficient internal energy into the atmosphere to vaporize an ocean mass of water, assuming the ocean is relatively evenly distributed over the surface. 
However, this is an upper limit because the atmosphere would also radiate a portion of its internal energy outwards to space, reducing the energy available for ocean vaporization. Approximately half of the atmospheric energy should be radiated to space, but it can vary depending on atmospheric composition and dynamics; for cooler steam atmospheres more energy is radiated to space \citep{Zahnle2020} whereas vapor saturated silicate atmospheres would be hotter at the bottom than the top and radiate more energy downwards. It should also be noted that thermal radiation is not the only mechanism to transfer energy from the atmosphere to the ocean; as the silicate vapor cools, molten silicate droplets falling into the ocean transfer latent heat that can contribute to ocean vaporization. Because the full dynamics and compositional evolution of the post-impact atmosphere are beyond the scope of this work, for simplicity we report the total internal energy deposited into the post-impact silicate atmosphere. 

\begin{figure}[t!]
\centering
\includegraphics[width=1.0\textwidth,trim={0 0cm 0 0cm},clip]{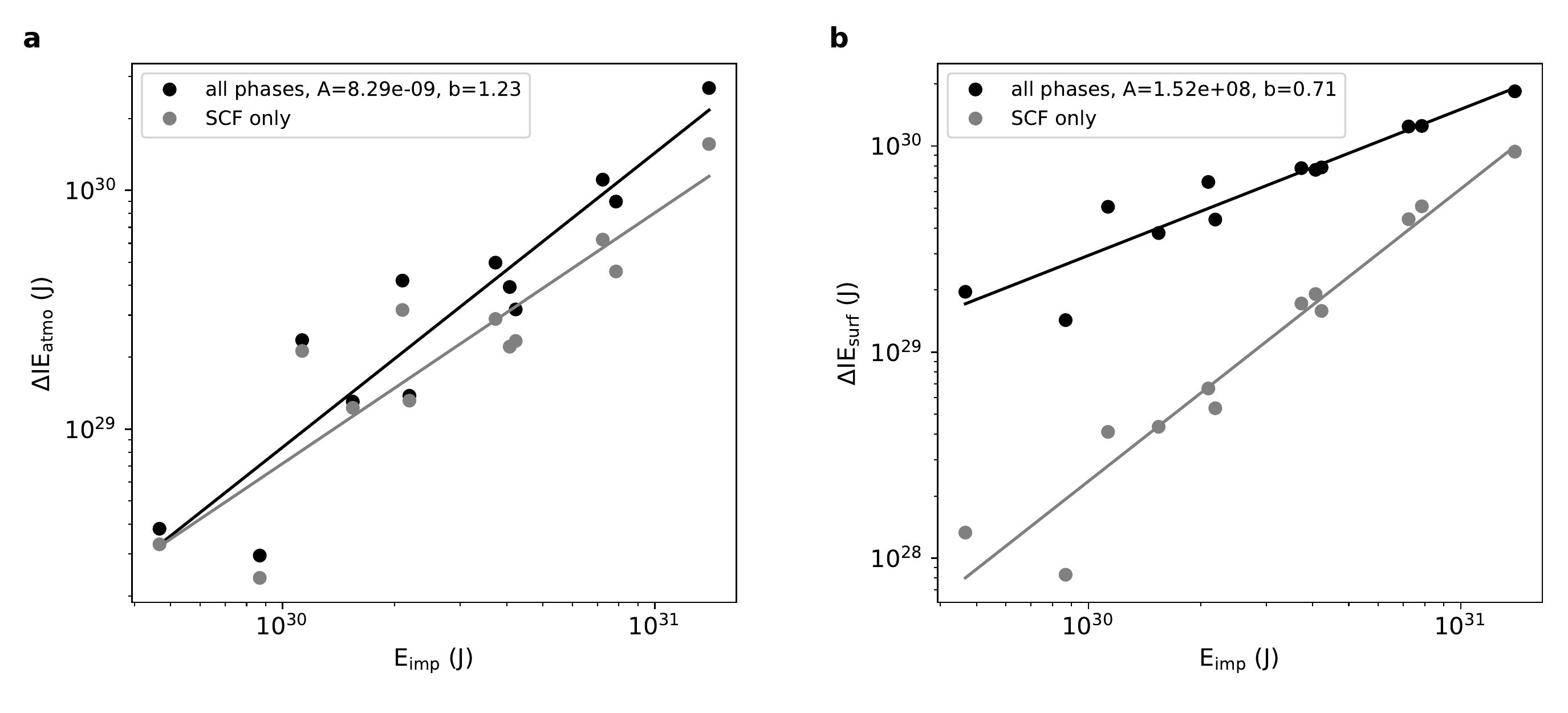}
\caption{(a) Energy increase in silicate atmospheric particles for $\theta$=45$^\circ$ simulations. (b) Energy increase in silicate surface particles for $\theta$=45$^\circ$ simulations. The energy increase for only the supercritical fluid (SCF) component of the atmosphere and surface is shown in grey. The values for the best fit to a power law ($A e^b$) are listed.  See Table \ref{tbl:energy_part}}
\label{fig:Eatmo}
\end{figure}

We extrapolate our results to determine the minimum impact size necessary to vaporize an ocean mass of water. Figure \ref{fig:Eatmo}a shows how the amount of energy deposited into the atmosphere varies with impact energy, plotted only for simulations with $\theta=45^\circ$ (because of the strong dependence on impact angle). For $\theta=45^{\circ}$, our results fit the power law $\Delta IE_{atmo} = 8.29\times10^{-9} (E_{imp})^{1.23}$, where $\Delta IE_{atmo}$ is the internal energy increase in the silicate atmosphere in Joules and $E_{imp}$ is the impact energy in Joules. This suggests that an impactor of mass $8.7 \times 10^{20}$ kg ($D \sim 710$ km) is required to vaporize one ocean mass of water ($5\times10^{27}$ J) if impacting at $45^\circ$ and 16.8 km/s (1.5 times Earth's escape velocity), assuming all of the internal energy increase in the forsterite atmosphere is applied towards water vaporization. If half of the energy is radiated to space then a $1.5 \times 10^{21}$ kg ($D \sim 860$ km) object would be required. 

For comparison, we also plot the amount of energy deposited in the surface layer (Figure \ref{fig:Eatmo}b), which is greater than the energy deposited in the atmosphere. Based on the fit in Figure \ref{fig:Eatmo}b, a $2.18 \times 10^{19}$ kg ($D \sim 210$ km) object impacting at 16.8 km/s could vaporize one ocean mass of water using the energy deposited in the outer layer of the planet. At this scale the impact energy deposited in the surface is expected to be localized to the region of impact (as shown in Section \ref{sec:global_melting}), making the estimate based on atmospheric energy potentially more applicable for global ocean vaporization. 
However, if the impact energy deposited into the surface is efficiently transported to the atmosphere via thermal radiation and re-radiated downwards then it could contribute to ocean vaporization. The cooling timescale of shallow magma oceans is potentially as low as hundreds of years \citep{Rubie2003}, although it could be much longer if a quenched crust forms (which may be likely for non-global events due to the rapid timescale of silicate atmosphere condensation). A 210 km diameter impactor providing sufficient energy to the surface layer to vaporize an ocean (5 $\times$ 10$^27$ J) results in approximately 300 W/m$^2$ of heating power if the magma pond cools over 1000 years and re-distributes its energy over the Earth's surface. This is greater than the 150 W/m$^2$ cooling rate of a steam atmosphere \citep{Sleep1989}, although it should be noted that not all of the heat radiated from the magma pond would be absorbed by the atmosphere, and at least half of the energy input into the atmosphere is likely radiated away into space (depending on the opacity profile of the atmosphere). The heat deposited in the surface layer therefore also likely contributes to ocean vaporization, but the magnitude of its contribution depends on the magma pond solidification timescale and the absorption and re-radiation of that heat in the atmosphere, a quantitative treatment of which has not been studied in detail and is beyond the scope of this work. 
Figure \ref{fig:Eatmo} also shows the internal energy increase in only the super-critical fluid component of the atmosphere and surface; the post-impact atmosphere is primarily composed of super-critical fluid with the remaining mass primarily vapor, whereas the liquid/solid surface layer contains only a small fraction of super-critical fluid (see also Table \ref{tbl:fors_part}).

\subsubsection{Global melting}\label{sec:global_melting}

\begin{figure}[t!]
\centering
\includegraphics[width=1.0\textwidth,trim={1cm 0.5cm 1cm 0.5cm},clip]{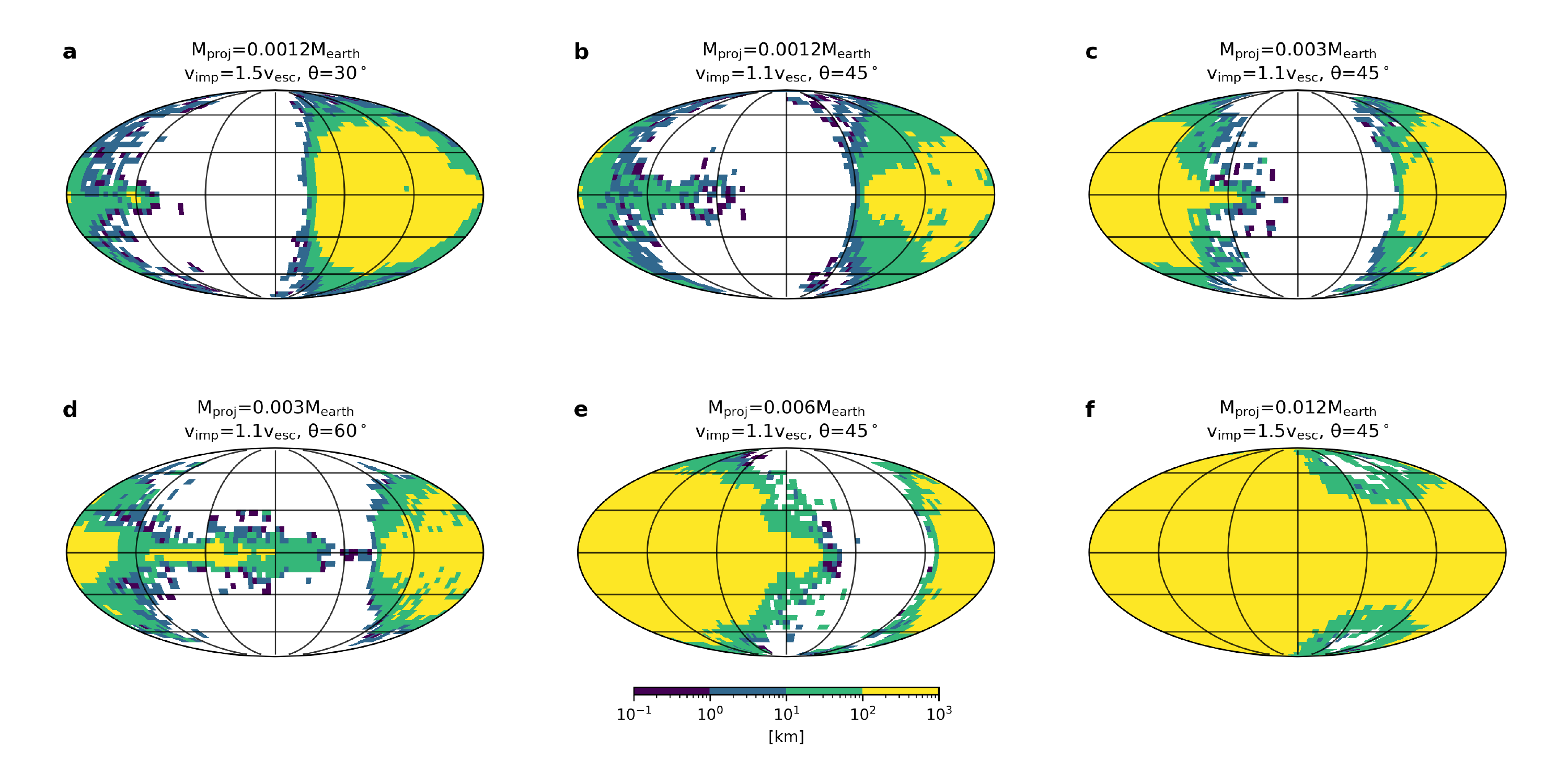}
\caption{Distribution of melt thickness 24 hrs post-impact for several example simulations. In the plotted coordinates the impact direction is east along the equator. The longitude of the impact point 24 hrs post-impact varies based on the planetary rotation induced by the impact.}
\label{fig:surface_melt}
\end{figure}

\begin{figure}[h!]
\centering
\includegraphics[width=0.9\textwidth,trim={0 0cm 0 0.5cm},clip]{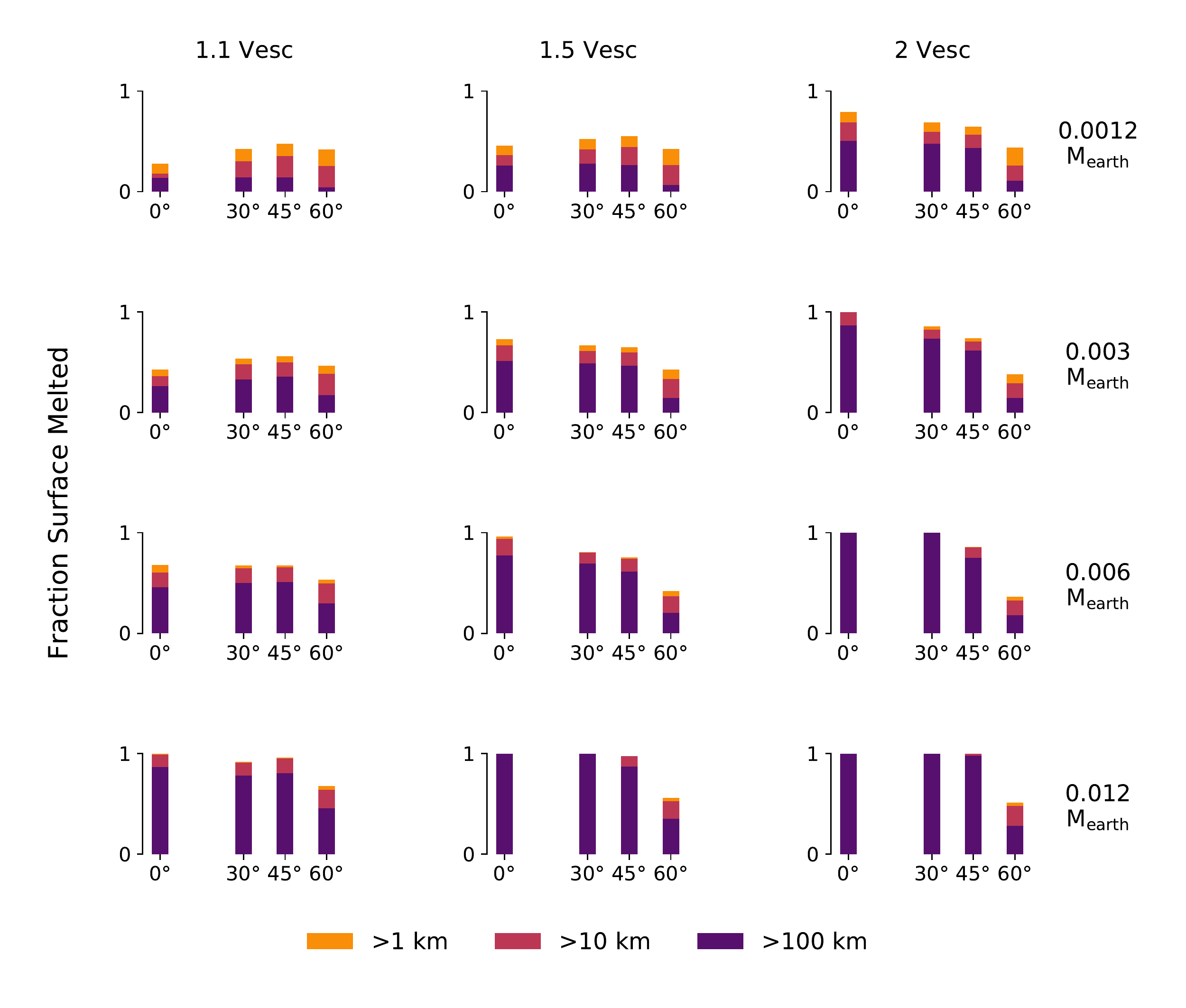}
\caption{Bar charts showing the fraction of surface covered $>$ 100, 10, or 1 km of melt after 24 hrs of simulation time. Charts are distributed in columns and rows according to the projectile velocity and mass. The x-axis in each bar chart shows runs of different impact angle (0$^\circ$ = head-on collision). }
\label{fig:stats_surfacemelt}
\end{figure}

To estimate sterilization from surface melting, we compute the equivalent melt thickness over a surface grid based on the degree of melting in surface particles 24 hours post-impact. For this calculation, super-critical fluid particles (only those in the surface) were counted as 100\% liquid melt. The distribution of post-impact surface melt is shown for several example simulations in Figure \ref{fig:surface_melt}, and the fraction of surface area covered in melt of various depths for all simulations is shown in Figure \ref{fig:stats_surfacemelt}. For smaller impacts ($M_{imp}$ $<$ 0.003 $M_{Earth}$), the melt is mostly concentrated in the impacted region, with additional melting downrange of the impacted region for more oblique impacts (Figure \ref{fig:surface_melt}a-d). For larger impacts, the majority of the surface is covered in melt over 100s of meters or 100s of kilometers thick (Figure \ref{fig:surface_melt}e,f). 

Our results imply that only the largest impacts cause sterilization via global surface melting, however, even the largest impactors we examined only partially melt the surface if they are oblique (60$^\circ$) (Figure \ref{fig:stats_surfacemelt}). 
It should be noted that this calculation excludes the effects of silicate condensation, and the post-impact rock-vapor atmosphere would eventually condense over the surface and increase the global melt thickness. Based on the mass of forsterite in the atmosphere (Table \ref{tbl:fors_part}), the condensed forsterite atmosphere averaged globally over the surface would be over 1 km thick for almost all simulations (Figure \ref{fig:stats_surfacemelt2}). 

We also show how the mass of the forsterite melt and atmosphere vary with impact energy (Figure \ref{fig:Mmelt}). The mass of the melt pool relative to the deposited iron can affect post-impact melt-atmosphere equilibration \citep[\eg,][]{Itcovitz2021}.

\subsection{Reducing environment}

\begin{figure}[h!]
\centering
\includegraphics[width=0.9\textwidth,trim={0 0cm 0 0.5cm},clip]{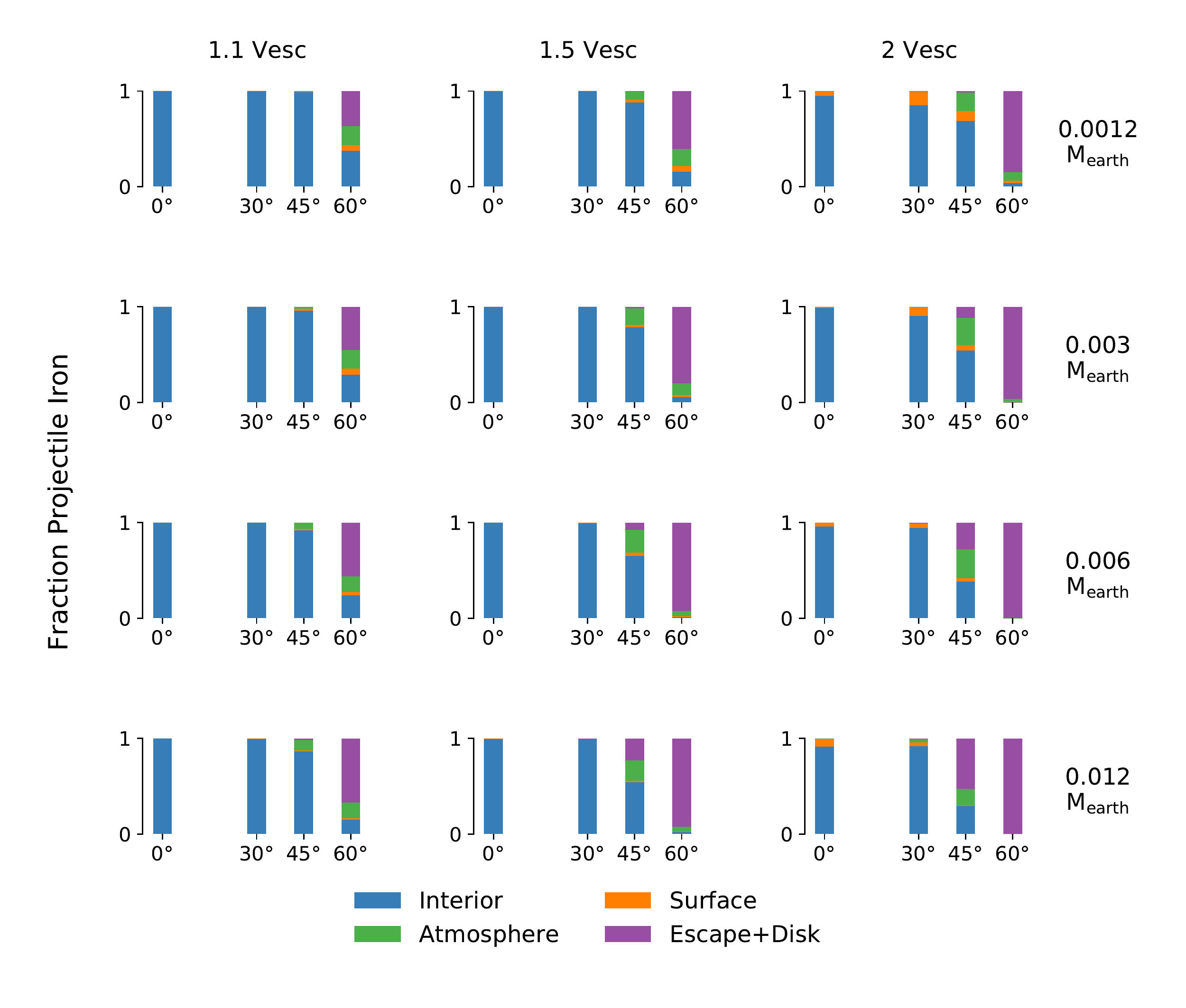}
\caption{Bar charts showing the distribution of projectile iron 24 hours post-impact. Projectile iron is divided into escaping material, atmospheric material, surface material ($P$ $>$ 10 GPa and $\rho$ $<$ 1000 kg/m$^3$) and interior material (anything below the surface material). Charts are distributed in columns and rows according to the projectile velocity and mass. The x-axis in each bar chart shows runs of different impact angle (0$^\circ$ = head-on collision). See Table \ref{tbl:iron_distribution}.}
\label{fig:iron_distribution}
\end{figure}

For each simulation, we quantify the amount of iron deposited into the interior, surface, atmosphere, and ejected from the system after 24 hours of simulation time, as shown in Figure \ref{fig:iron_distribution}. 
The post-impact distribution of projectile iron depends strongly on the impact angle. In more direct impacts ($\theta = 30^{\circ}$), the projectile iron burrows deep into the Earth's mantle where overlying mantle material prevents its expulsion to the surface during post-impact decompression and rebound, sequestering almost all of the projectile iron within the mantle. For more oblique impacts ($\theta = 60^{\circ}$), a significant amount of iron is ejected from the system, a result of the hit-and-run nature of these impacts, which is more pronounced for higher impact velocities. The amount of iron deposited onto the surface and into the post-impact atmosphere is higher for more oblique impact angles, and also weakly increases with impactor mass and velocity.  

The reducing potential of the post-impact atmosphere depends on the amount of iron delivered to reservoirs where it can reduce significant quantities of H$_2$O. \citet{Benner2020} and \citet{Zahnle2020} suggest that 1$-$3 ocean masses of water must be reduced to produce an atmosphere sufficiently rich in H$_2$ and CH$_4$ to generate the necessary RNA precursor compounds to initiate and sustain RNA-based life. However, not all projectile iron delivered to the Earth during an impact is expected to interact with and reduce surface water, a point emphasized by \citet{Zahnle2020}.
Iron deposited in the mantle during the impact would likely have had insufficient exposure to react with surface H$_2$O or CO$_2$, and a significant fraction of the large iron fragments input into the mantle likely sink to the core either as descending diapirs or propagating dikes \citep{Tonks1992,Rubie2015}. Similarly, much of the iron deposited onto the surface may be buried beneath rock ejecta or deposited in regions where the post-impact crust is molten. Comparing the post-impact distribution of surface iron for several example simulations (Figure \ref{fig:surface_iron}) with the corresponding post-impact surface melt distribution (Figure \ref{fig:surface_melt}) shows that most of the iron deposited on the surface is located in highly melted regions where it may sink into the lower crust and be sequestered before reacting with surface water. The material sequestered in the mantle and upper crust may still be important for contributing reduced gases to the atmosphere on longer geologic timescales. However, because of large uncertainties in the processing timescale of such material, it is unclear if sequestered iron would release sufficient reducing gases to augment the post-impact reducing atmosphere; the post impact reducing environment persists for an upper limit of 10-100 Myr based on the assumption that all projectile iron (in a maximum-HSE scale impact) is deposited in the atmosphere \citep{Benner2020,Zahnle2020}. An important caveat is that if the post-impact melt pool is allowed to equilibrate with the post-impact atmosphere then iron deposited into the surface melt could significantly alter the redox state of the post-impact atmosphere; this process is beyond the scope of this work but is examined in detail in \citet{Itcovitz2021} (see Section \ref{sec:reducingpotential} for further discussion).

The primary source of reducing iron may therefore be iron that is vaporized during the impact and retained in the post-impact atmosphere. The post-impact atmosphere is a mixture of iron, rock, and surface water that were vaporized during the impact into a gas and/or supercritical fluid state. Iron deposited in the post-impact atmosphere should therefore react relatively quickly with water vapor/supercritical fluid, depending on the oxygen fugacity of the system \citep[\eg,][]{Choudhry2014}. 
Atmospheric iron would eventually condense and rain out in small droplets, but could react with water prior, during, or after rain-out.
The total mass of post-impact atmospheric iron provides an upper limit on the amount of water that could be reduced because a fraction of the iron deposited in the atmosphere would rain out over the molten portion of the planetary surface and potentially be sequestered before reacting with water. However, computing the exact fraction of iron that can reduce H$_2$O before raining out over molten surfaces is beyond the scope of this work. Larger impacts can result in near total surface melting, which could efficiently sequester condensing atmospheric iron, however, such impacts also produce more massive and hotter post-impact atmospheres that may allow for longer residence times of atmospheric iron prior to condensation, and therefore more efficient reactivity with vaporized water. 
For simplicity we neglect such effects and assume that all iron deposited in the atmosphere is available to reduce surface water.

\begin{figure}[h!]
\centering
\includegraphics[width=0.9\textwidth,trim={0 0cm 0 0.5cm},clip]{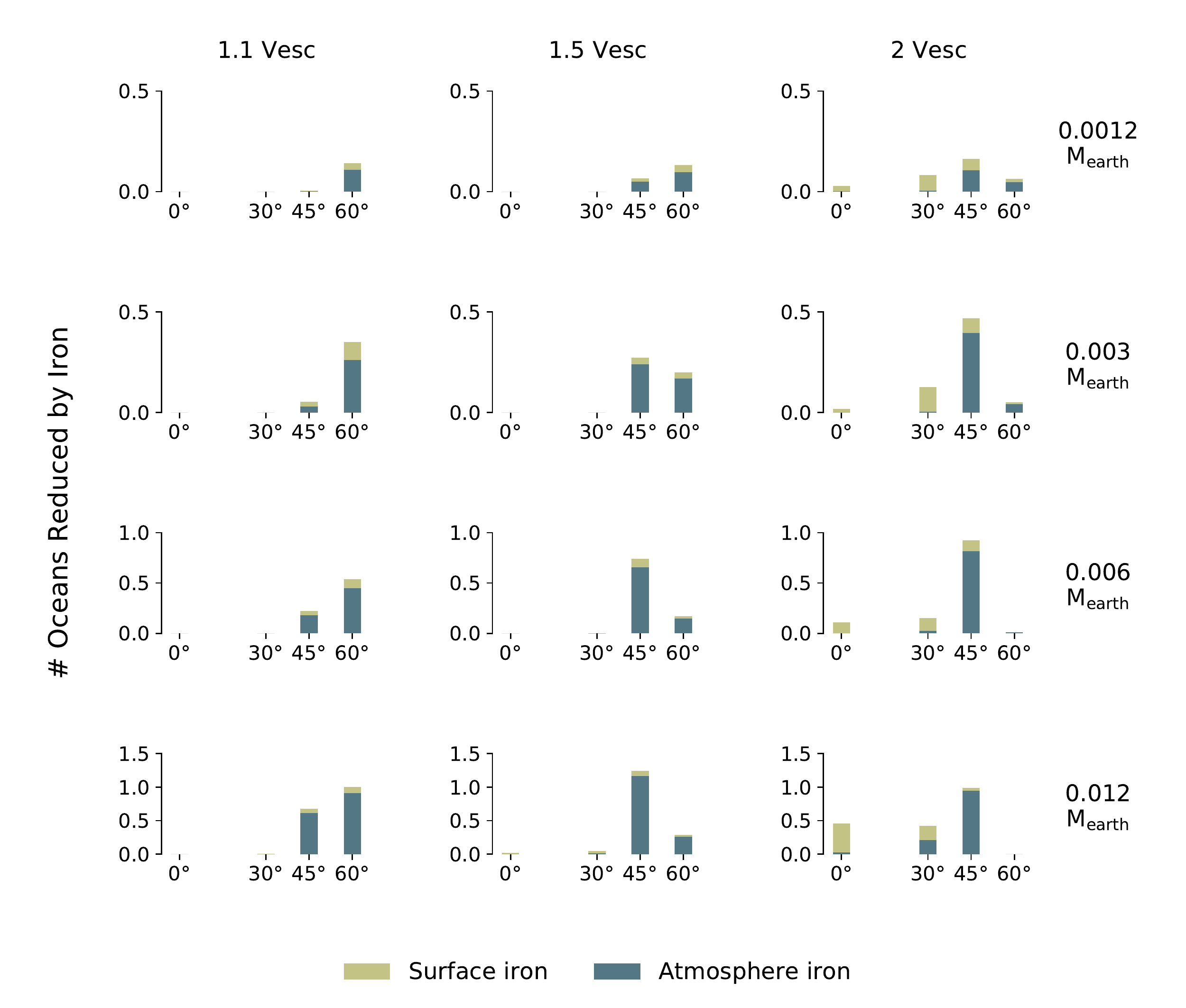}
\caption{Bar charts showing the number of oceans that could be reduced by the iron in the atmosphere and deposited in the surface layer. We assume 10$^{22}$ kg of iron can reduce 2.3 oceans of water. Charts are distributed in columns and rows according to the projectile velocity and mass. The x-axis in each bar chart shows runs of different impact angle (0$^\circ$ = head-on collision). Note that the y-axis scale changes depending on the projectile mass. }
\label{fig:reduced_oceans}
\end{figure}

The number of oceans that could be reduced by the projectile iron deposited in the post-impact atmosphere is shown in Figure \ref{fig:reduced_oceans}. In our simulations, only a few high velocity and high mass impacts at $\theta=$45$^{\circ}$ delivered sufficient iron to the atmosphere to reduce an ocean mass of water. 
Impacts of mass 0.003 $M_{Earth}$ delivered sufficient iron to the atmosphere to reduce a maximum of $\sim$ 0.5 ocean masses of water. 
Figure \ref{fig:reduced_oceans} also shows the amount of iron deposited in the surface layer, in case this iron is able to react with surface or near-surface water before being sequestered in the solidifying crust. In general the amount of iron deposited in the surface layer is much less than that deposited in the atmosphere, except for in head-on or highly incident collisions ($\theta$ = 0$^\circ$ or 30$^\circ$) where more iron is deposited in the surface layer than the atmosphere. 
Considering the combined effect of atmospheric and surface iron, a maximum of $\sim$ 1.25 ocean masses of water is reduced by impact-delivered iron.

\section{Discussion}

\subsection{Reducing potential of post-impact atmospheres}\label{sec:reducingpotential}

Our results suggest that large impacts during late accretion were inefficient at delivering reactive iron to the surface/atmosphere. Significant fractions of projectile iron are either buried in the crust, sequestered in the mantle (where large fragments likely sink to the core), or ejected from the system and therefore unable to reduce surface/atmospheric water and significantly alter the atmospheric redox state. Furthermore, a large fraction of the iron delivered to the atmosphere is expected to condense and rain out over molten surfaces (depending on the fugacity), where it could become sequestered in the solidifying crust and upper mantle before reacting with surface/atmospheric water to produce hydrogen. 
Limited interaction between projectile iron and surface water is problematic to the view that the largest of the late accretion impactors could have completely reduced Earth's surface water and CO$_2$ to generate a transient reducing atmosphere. The canonical scenario for generating a high-reducing post-impact atmosphere relies on most of late accretion being concentrated in a single large impactor 2500-3000 km in diameter, containing sufficient iron ($\sim10^{22}$ kg) to reduce 2.3 ocean masses of water \citep{Benner2020,Zahnle2020}. Almost all of this iron would be required to reduce Earth's surface water if the early oceans were two times more massive than today (massive early oceans are plausible given the lower expected water retention of Earth's early mantle \citep{Dong2021}).
While prior works have assumed that all of the iron delivered in late accretion impacts is immediately available to interact with surface water, our results suggest that only 20\% or less of the projectile iron is delivered to the post-impact atmosphere, and much of the iron could be buried prior to reacting with H$_2$O.

Despite inefficient retention of projectile iron at the surface, several scenarios could still allow for the complete reduction of surface water. First, inefficient retention of projectile iron in the Earth's mantle could necessitate a late accretion mass 2-5 times more massive than normally assumed to explain the observed excess of mantle HSEs \citep{Marchi2017}. Delivery of more iron by a more massive late accretion impactor could feasibly have still reduced 1-3 ocean masses of water, even with inefficient delivery of iron to the post-impact atmosphere. This scenario is also feasible without increasing the projectile mass if the projectile core fraction is increased, which is plausible given the diversity of planetesimals predicted by \citet{Carter2015}. 

Second, while a significant portion of projectile iron may have been sequestered in the crust and upper mantle, an efficient mechanism to oxidize that iron and release reducing gases may have bolstered the reducing power of the post-impact atmosphere. On short timescales, if most of the iron deposited into the crust and mantle remained in the impact-generated melt pool it could react to produce a reduced melt phase; if the reduced impact-generated melt was able to equilibrate with the post-impact atmosphere the atmosphere would become more reduced than expected based solely on the amount of iron deposited into the atmosphere \citep{Zahnle2020,Itcovitz2021}. In particular, such equilibration of the atmosphere with a melt phase of peridotite-like composition (which is likely based on high-fraction melt dominating the melt composition) that has been reduced by iron deposited in the target interior can result in a post-impact atmosphere nearly as reduced as would be expected if all of the projectile iron was deposited directly into the atmosphere \citep{Itcovitz2021}. However, this effect diminishes and the post-impact atmosphere becomes less reducing if iron is deposited into portions of the mantle or melt pool that do not participate in post-impact melt-atmosphere equilibration \citep{Itcovitz2021}. If iron rapidly pools at the base of the melt pool, then the timescale of chemical equilibration is likely several orders of magnitude longer than the timescale of magma pond solidification; however, if the projectile iron is emulsified into smaller droplets with longer settling times then significant equilibration could occur within the magma pond \citep{Rubie2003}. The scale of fragmentation of the projectile core is much less than the particle size in our simulations, but \citet{Genda2017} estimated in similar simulations that $\sim$ 50\% of the projectile core would be fragmented into 10 meter average-size blobs based on the strain rate experienced during the pressure release process. After the impact, further passive settling of these metal masses reduce them to decimeter scale droplets through turbulent entrainment and Rayleigh Taylor and Kelvin-Helmholtz instabilities as they sink through the impact melt \citep{Rubie2003,Dahl2010,Deguen2014}. Larger fragments of inertially confined iron that penetrate into the mantle can also break into smaller fragments due to pancaking and stretching during the impact \citep{Kendall2016}. 

Equilibration between the melt and atmosphere could also be inhibited by the formation of a surface fusion crust or cooling of the melt below the rheological transition for vigorous convection. 
However, even if iron deposited into the crust and mantle does not immediately equilibrate with the post-impact atmosphere, sequestered iron could still influence the redox state of volcanic gases released over geological timescales. Unoxidized iron in the crust and mantle could generate reduced gases such as H$_2$ and CH$_4$ by reacting with water or CO$_2$, respectively. Enhanced production of such gases from iron embedded in the crust and mantle could have lasted for 10s of Myr or more \citep{Zahnle2020} because more reduced magmas can sustain long-term H$_2$ production \citep{Liggins2020}, potentially maintaining a reducing atmosphere in the aftermath of a large impact. Exactly how efficient this process would be involves many complex factors relating to the timescale of iron oxidation in the crust and mantle coupled to rates of volcanic out-gassing, and is beyond the scope of this work. We also again emphasize that a significant fraction of the large iron fragments sequestered in the mantle are expected to sink to the core, and the retention of significant projectile iron in the mantle depends on the efficiency of iron fragmentation and emulsification during the impact and subsequent melt pond solidification.

Third, ejected material could bolster post-impact reducing atmospheres if it re-impacts the Earth. Ejected material would be deposited in a ring of debris sharing a similar orbit to Earth \citep{Genda2017b,Marchi2017}, and numerical estimates of ejecta reaccretion suggest that 20$-$30\% of ejected iron fragments re-impact the Earth within 10s of Myr for giant impacts \citep{Jackson2012,Bottke2015}, although this timescale may be orders of magnitude less for smaller impacts. In oblique $60^\circ$ impacts where almost all projectile iron is ejected, 20\% of the ejected iron is comparable in mass to the amount of iron deposited in the atmosphere during more nominal $45^\circ$ impacts, suggesting that oblique $60^\circ$ impacts could deliver a similar amount of reducing power even though the majority of iron initially escapes (Figure \ref{fig:iron_distribution}). The reaccreted iron may actually be more likely to react with surface/atmospheric water than post-impact atmospheric iron because reaccreting iron is delivered in smaller impacts that melt less of the surface, increasing the likelihood that reaccreting iron reacts with surface water instead of being sequestered in the impact melt. However, iron deposited during reaccretion is distributed over 10s of kyr to 10s of Myr, allowing for hydrogen escape and oxidation between impacts, which lowers the maximum reducing power when compared to a comparable mass of iron injected directly into the atmosphere in a single impact event. This effect may be mitigated if the ejected iron consists of a few large fragments, as expected during hit-and-run type collisions, allowing large masses of iron to be delivered in single reaccretion events.

Even if the largest late accretion impactor did not completely reduce all of Earth's surface water, there are pathways to an RNA world that could arise from smaller amounts of iron delivered to the surface. \citet{Benner2019} suggest that mid-sized impactors $10^{20}-10^{23}$ kg would still generate transient weakly-reduced atmospheres that could produce reduced species such as nucleobase precursors and ammonia. Because less iron is delivered in such impact events the atmosphere would return to redox equilibrium with the mantle more quickly, which in some ways can be advantageous because it allows the prebiotic system to utilize the reduced precursors produced in the post-impact atmosphere while also exploiting chemical pathways that rely on an atmosphere closer to redox equilibrium with an FMQ mantle \citep{Benner2019}. The combination of volcanic SO$_2$ gas, an FMQ mantle, and reduced precursors from a transient post-impact atmosphere, could provide a direct pathway for the emergence of an RNA world in the Hadean \citep{Benner2019}.

\subsection{Continuity of habitability during late accretion}

If life originated in a post-impact reducing environment, there must have been no subsequent planet-wide sterilization events. This may have been difficult because the impacts large enough to deliver sufficient iron to form post-impact reducing atmospheres are generally larger than the minimum size impact that could sterilize the Earth's surface. Due to the decreasing impactor size over time, large impacts capable of reducing all of Earth's surface water could have been followed by smaller impacts capable of sterilizing the Earth's surface. The likelihood of subsequent sterilization depends on the size distribution and timing of late accretion impacts, and the minimum size impact required to sterilize the planet.

However, in general we find that the size of impact required to sterilize the planet is larger than prior estimates. For sterilization via global surface melting, prior estimates suggest a 1300-2000 km diameter impactor is required to completely melt the surface \citep{Marchi2014,Svetsov2007}, whereas we find that 2000 km diameter impacts can leave large fractions ($>$ 50\%) of the surface unmelted, depending on the impact parameters, with the threshold for complete surface melting lying somewhere between 2000 and 2700 km diameter impacts. For sterilization via ocean vaporization, prior studies assumed a 350-450 km diameter impact could vaporize an ocean mass of water \citep{Sleep1989,Zahnle2020}, whereas we estimate a 740 km diameter impactor would be required based on the amount of energy in the post-impact atmosphere (for $\theta=45^\circ$ and $v_{imp}=1.5v_{esc}$). The 740 km diameter threshold for ocean vaporization is also only a lower limit and radiation of atmospheric energy to space would necessitate a larger impact. However, it should be noted that smaller impacts that only partially vaporize the early ocean could still be sterilizing events if they annihilate the photic zone and turn early oceans into an uninhabitable hot dense brine \citep{Sleep1989}. Additionally, if heat is effectively and rapidly transferred into the atmosphere from the impact melt pool and re-radiated downwards then impacts as small as 200 km diameter could also be ocean vaporizing events. Alternatively, it is unclear if complete ocean vaporizing impacts would be completely sterilizing events, because although radiative heat from the post-impact atmosphere could boil away surface water, lifeforms could persist in subsurface habitable environments that were either pre-existing or buried during the excessive ground shaking that accompanies large impact events.

The need for a larger impact to sterilize the Earth's surface may allow for the origin of life in the aftermath of a mid to large sized impact without sterilization from subsequent impact events. Sterilization is only completely ensured by an impact $>$2000 km in diameter, and it is unclear if any such impacts would have post-dated a 3000 km diameter impact capable of generating a strong reducing atmosphere \citep{Benner2020}. An impactor able to deliver sufficient iron to catalyze a post-impact origin of life may therefore have also been the last planet-wide sterilizing impact \citep{Benner2020}. 
Sterilization subsequent to the origin of life becomes more likely if an ocean-vaporizing impact is sufficient, however, some dynamic models of accretion suggest that even an 800 km diameter ocean-vaporizing impact is unlikely to occur after the delivery of the excess HSE budget to the Earth's mantle \citep{Brasser2020}. The chance of subsequent sterilization also decreases if life was derived from a more weakly reducing atmosphere in the aftermath of a smaller $<$ 2000 km diameter impact \citep[\eg][]{Benner2019}, because the largest subsequent impact would likely also have been smaller.

\subsection{Model limitations and future work}

The results of our study, for both planetary sterilization and iron delivery, contain a degree of uncertainty related to our numerical modeling. In particular, we ignore material strength and treat the target planet and impactor as fluid bodies. While this is a commonly used approximation in models of large impacts, recent work has shown that material strength can affect the dynamics of basin scale impacts \citep{Emsenhuber2018}. 
Inclusion of material strength results in more localized deposition of the impactor's kinetic energy, which can reduce the overall surface area covered in post-impact melt. This suggests that the minimum size impact required to globally melt the surface would be even higher than predicted in our fluid simulations. Material strength may also increase the size of impact required to vaporize oceans, because more projectile kinetic energy is required to produce a given amount of plastic work prior to decompression and vaporization, which can reduce the overall fraction of the impact kinetic energy that is eventually deposited into the atmosphere. Material strength would also affect the post-impact iron distribution, because more iron would be retained in the upper mantle if the planet is not treated as a fluid. The full effect of material strength on these aspects of the simulations will be examined in future work. Artificial viscosity can also affect material mixing in SPH simulations, and future work will also test varying the artificial viscosity parameters. Overall, due to the high velocity and scale of the expected impacts, in addition to the relatively warm state of the early Earth, the fluid calculations presented in this study provide reasonable estimates of the post-impact environment of large impacts on the early Earth. 

An additional source of uncertainty is the initial thermal state of the Earth. We assumed a surface temperature of $\sim$ 1900 K due to the expected warm state of the early Earth's mantle and our inability to properly resolve the cool outer crust in resolution-limited 3D simulations. A cooler initial thermal profile, or the inclusion of a cold lithosphere, could affect our results by requiring more impact deposited energy to achieve melting. However, a change in the surface temperature by a few hundred degrees is unlikely to strongly affect our results because it is a small effect compared to the large amount of impact delivered energy. This can be tested in future work that accounts for varying initial temperature profiles. 

We also note that our analysis rests on the large scale distribution of energies in 3D simulations that utilize indivisible particles $\sim$150-250 km in diameter; we therefore do not resolve any crust or atmospheric mixing. Future work at higher resolution could provide more insight into finer scale processes.
Our 3D simulations were also computationally limited to examine projectiles only as small as 1500 km in diameter. Modeling smaller impacts with projectiles as small as 200 or 500 km in diameter will also be the focus of future work as computational resources improve and SPH codes become more efficient, providing improved constraints on the threshold for sterilizing impacts in addition to the more localized effects of such impacts.

\section{Conclusions}

We find that late accretion impacts are generally less sterilizing than previously assumed but also deliver less metallic iron to the post-impact atmosphere, making the development of post-impact highly-reducing atmospheres more difficult. The effects of late accretion impacts on early Earth habitability depend strongly on impact parameters. In general, our simulations show that most impact-delivered iron would be sequestered into the mantle. Only the largest impacts (3400 km diameter) deliver sufficient iron to the post-impact surface and atmosphere to reduce an entire ocean mass of water, and these impacts also melt most of the planetary surface, which might entrain delivered iron prior to its interaction with water. Post-impact atmospheres may therefore be less reducing than expected based on the total mass of iron in the impactor. However, even weakly reducing atmospheres may still open pathways to an RNA world \citep{Benner2019}, and interactions between the atmosphere and the impact-generated melt phase in the aftermath of the impact could act to restore reducing power to the atmosphere \citep{Itcovitz2021}. If life did originate in a post-impact environment, its persistence depended on the likelihood of a subsequent sterilizing impact. We find that sterilizing impact events require larger projectiles than previously assumed. Our simulations suggest that impactors of mass 0.006 $M_{Earth}$ can sterilize the Earth's surface through global melting, and impactors at least 700 km in diameter could have vaporized Earth's early oceans. This reduces the relative likelihood of a sterilizing impact event, however, several such impacts are still expected to have occurred during late accretion. The feasibility of generating life in a post-impact environment thus involves a complex balance between requiring an impact sufficiently large to deliver large amounts of reactive iron to the surface, but not so large that subsequent sterilizing events are highly probable. This is further complicated by the stochastic nature of planetary accretion that determines the timing and likelihood impacts, because varying impact parameters can strongly affect iron delivery and planetary sterilization; for example, highly oblique impacts are more likely to leave portions of the surface unmelted and habitable. Further work on the timing and size distribution of late accretion impacts, the consequences of such impacts in terms of planetary sterilization and iron delivery, and the processing of iron deposited in the crust and upper mantle, will help us to better understand the likelihood of an origin of life that is catalyzed by impact-delivered iron.

\section*{Acknowledgements} 
STS and RIC are supported by Simons Foundation Grant \#554203. We thank Norm Sleep and an anonymous reviewer for comments that helped to improve this manuscript. We thank Jonathan Itcovitz, Oliver Shorttle, Auriol Rae, David Catling, Nick Wogan, and Phil Carter for useful discussions. We also thank Kevin Zahnle for comments on a draft of this manuscript. The modified version of GADGET-2 for planetary impacts is available from the supplement of \cite{Cuk2012b} and the equation of state tables from \cite{stewart2019zenodo-forsterite,stewart2020zenodo-ironalloy}.

\bibliography{references.bib}

\begin{thebibliography}{}
\expandafter\ifx\csname natexlab\endcsname\relax\def\natexlab#1{#1}\fi
\providecommand{\url}[1]{\href{#1}{#1}}
\providecommand{\dodoi}[1]{doi:~\href{http://doi.org/#1}{\nolinkurl{#1}}}
\providecommand{\doeprint}[1]{\href{http://ascl.net/#1}{\nolinkurl{http://ascl.net/#1}}}
\providecommand{\doarXiv}[1]{\href{https://arxiv.org/abs/#1}{\nolinkurl{https://arxiv.org/abs/#1}}}

\bibitem[{Abelson(1966)}]{abelson1966chemical}
Abelson, P.~H. 1966, Proceedings of the National Academy of Sciences of the
  United States of America, 55, 1365

\bibitem[{Abramov \& Mojzsis(2009)}]{Abramov2009}
Abramov, O., \& Mojzsis, S.~J. 2009, Nature, 459, 419,
  \dodoi{10.1038/nature08015}

\bibitem[{Abramov {et~al.}(2012)Abramov, Wong, \& Kring}]{Abramov2012}
Abramov, O., Wong, S.~M., \& Kring, D.~A. 2012, Icarus, 218, 906,
  \dodoi{10.1016/J.ICARUS.2011.12.022}

\bibitem[{Barr \& Citron(2011)}]{Barr2011}
Barr, A.~C., \& Citron, R.~I. 2011, Icarus, 211, 913

\bibitem[{Benner {et~al.}(2019)Benner, Kim, \& Biondi}]{Benner2019}
Benner, S.~A., Kim, H.~J., \& Biondi, E. 2019, Life, 9, 1,
  \dodoi{10.3390/life9040084}

\bibitem[{Benner {et~al.}(2020)Benner, Bell, Biondi, Brasser, Carell, Kim,
  Mojzsis, Omran, Pasek, \& Trail}]{Benner2020}
Benner, S.~A., Bell, E.~A., Biondi, E., {et~al.} 2020, ChemSystemsChem, 2,
  \dodoi{10.1002/syst.201900035}

\bibitem[{Boehnke \& Harrison(2016)}]{boehnke2016illusory}
Boehnke, P., \& Harrison, T.~M. 2016, Proceedings of the National Academy of
  Sciences, 113, 10802, \dodoi{10.1073/pnas.1611535113}

\bibitem[{Bottke {et~al.}(2015)Bottke, Vokrouhlick{\'{y}}, Marchi, Swindle,
  Scott, Weirich, \& Levison}]{Bottke2015}
Bottke, W.~F., Vokrouhlick{\'{y}}, D., Marchi, S., {et~al.} 2015, Science, 348,
  321, \dodoi{10.1126/science.aaa0602}

\bibitem[{Bottke {et~al.}(2010)Bottke, Walker, Day, Nesvorny, \&
  Elkins-Tanton}]{Bottke2010}
Bottke, W.~F., Walker, R.~J., Day, J.~M., Nesvorny, D., \& Elkins-Tanton, L.
  2010, Science, 330, 1527, \dodoi{10.1126/science.1196874}

\bibitem[{Brasser {et~al.}(2016)Brasser, Mojzsis, Werner, Matsumura, \&
  Ida}]{Brasser2016}
Brasser, R., Mojzsis, S.~J., Werner, S.~C., Matsumura, S., \& Ida, S. 2016,
  Earth and Planetary Science Letters, 455, 85,
  \dodoi{10.1016/j.epsl.2016.09.013}

\bibitem[{Brasser {et~al.}(2020)Brasser, Werner, \& Mojzsis}]{Brasser2020}
Brasser, R., Werner, S.~C., \& Mojzsis, S.~J. 2020, Icarus, 338,
  \dodoi{10.1016/j.icarus.2019.113514}

\bibitem[{{Brian Tonks} \& {Jay Melosh}(1992)}]{Tonks1992}
{Brian Tonks}, W., \& {Jay Melosh}, H. 1992, Icarus, 100, 326,
  \dodoi{10.1016/0019-1035(92)90104-F}

\bibitem[{Canup \& Asphaug(2001)}]{Canup2001}
Canup, R.~M., \& Asphaug, E. 2001, Nature, 412, 708, \dodoi{10.1038/35089010}

\bibitem[{Carter {et~al.}(2015)Carter, Leinhardt, Elliott, Walter, \&
  Stewart}]{Carter2015}
Carter, P.~J., Leinhardt, Z.~M., Elliott, T., Walter, M.~J., \& Stewart, S.~T.
  2015, The Astrophysical Journal, 813, 72, \dodoi{10.1088/0004-637X/813/1/72}

\bibitem[{{Chou, C.-L.}(1978)}]{Chou1978}
{Chou, C.-L.} 1978, Proceedings of the 9th Lunar and Planetary Science
  Conference, 1, 219

\bibitem[{Choudhry {et~al.}(2014)Choudhry, Carvajal-Ortiz, Kallikragas, \&
  Svishchev}]{Choudhry2014}
Choudhry, K.~I., Carvajal-Ortiz, R.~A., Kallikragas, D.~T., \& Svishchev, I.~M.
  2014, Corrosion Science, 83, 226, \dodoi{10.1016/j.corsci.2014.02.019}

\bibitem[{{\'{C}}uk \& Stewart(2012)}]{Cuk2012b}
{\'{C}}uk, M., \& Stewart, S.~T. 2012, Science, 338, 1047,
  \dodoi{10.1126/science.1225542}

\bibitem[{Dahl \& Stevenson(2010)}]{Dahl2010}
Dahl, T.~W., \& Stevenson, D.~J. 2010, Earth and Planetary Science Letters,
  295, 177, \dodoi{10.1016/j.epsl.2010.03.038}

\bibitem[{Dale {et~al.}(2012)Dale, Burton, Greenwood, Gannoun, Wade, Wood, \&
  Pearson}]{Dale2012}
Dale, C.~W., Burton, K.~W., Greenwood, R.~C., {et~al.} 2012, Science, 335, 72,
  \dodoi{10.1126/science.1214967}

\bibitem[{Day {et~al.}(2007)Day, Pearson, \& Taylor}]{Day2007}
Day, J.~M., Pearson, D.~G., \& Taylor, L.~A. 2007, Science, 315, 217,
  \dodoi{10.1126/science.1133355}

\bibitem[{de~Vries {et~al.}(2016)de~Vries, Nimmo, Melosh, Jacobson, Morbidelli,
  \& Rubie}]{DeVries2016}
de~Vries, J., Nimmo, F., Melosh, H.~J., {et~al.} 2016, Progress in Earth and
  Planetary Science, 3, 1, \dodoi{10.1186/s40645-016-0083-8}

\bibitem[{Deguen {et~al.}(2014)Deguen, Landeau, \& Olson}]{Deguen2014}
Deguen, R., Landeau, M., \& Olson, P. 2014, {Turbulent metal-silicate mixing,
  fragmentation, and equilibration in magma oceans},
  \dodoi{10.1016/j.epsl.2014.02.007}

\bibitem[{Dong {et~al.}(2021)Dong, Fischer, Stixrude, \&
  Lithgow-bertelloni}]{Dong2021}
Dong, J., Fischer, R.~A., Stixrude, L.~P., \& Lithgow-bertelloni, C.~R. 2021,
  AGU Advances, 2, \dodoi{doi.org/10.1029/2020AV000323}

\bibitem[{Emsenhuber {et~al.}(2018)Emsenhuber, Jutzi, \& Benz}]{Emsenhuber2018}
Emsenhuber, A., Jutzi, M., \& Benz, W. 2018, Icarus, 301, 247,
  \dodoi{10.1016/J.ICARUS.2017.09.017}

\bibitem[{Genda {et~al.}(2017{\natexlab{a}})Genda, Brasser, \&
  Mojzsis}]{Genda2017}
Genda, H., Brasser, R., \& Mojzsis, S.~J. 2017{\natexlab{a}}, Earth and
  Planetary Science Letters, 480, 25, \dodoi{10.1016/j.epsl.2017.09.041}

\bibitem[{Genda {et~al.}(2017{\natexlab{b}})Genda, Iizuka, Sasaki, Ueno, \&
  Ikoma}]{Genda2017b}
Genda, H., Iizuka, T., Sasaki, T., Ueno, Y., \& Ikoma, M. 2017{\natexlab{b}},
  Earth and Planetary Science Letters, 470, 87,
  \dodoi{10.1016/j.epsl.2017.04.035}

\bibitem[{Gomes {et~al.}(2005)Gomes, Levison, Tsiganis, \&
  Morbidelli}]{Gomes2005}
Gomes, R., Levison, H.~F., Tsiganis, K., \& Morbidelli, A. 2005, Nature, 435,
  466, \dodoi{10.1038/nature03676}

\bibitem[{Hartmann(1975)}]{hartmann1975lunar}
Hartmann, W.~K. 1975, Icarus, 24, 181, \dodoi{10.1016/0019-1035(75)90095-0}

\bibitem[{Holland(1962)}]{Holland1962}
Holland, H.~D. 1962, in {Petrologic Studies} (Geological Society of America),
  \dodoi{10.1130/Petrologic.1962.447}

\bibitem[{Itcovitz {et~al.}(2021)Itcovitz, Rae, Shorttle, Citron, Stewart,
  Rimmer, \& Sinclair}]{Itcovitz2021}
Itcovitz, J.~P., Rae, A. S.~P., Shorttle, O., {et~al.} 2021, (in prep)

\bibitem[{Ivanov \& Melosh(2003)}]{ivanov2003impacts}
Ivanov, B., \& Melosh, H. 2003, Geology, 31, 869

\bibitem[{Jackson \& Wyatt(2012)}]{Jackson2012}
Jackson, A.~P., \& Wyatt, M.~C. 2012, Monthly Notices of the Royal Astronomical
  Society, 425, 657, \dodoi{10.1111/j.1365-2966.2012.21546.x}

\bibitem[{Kasting(2014)}]{Kasting2014}
Kasting, J.~F. 2014, Special Paper of the Geological Society of America, 504,
  19, \dodoi{10.1130/2014.2504(04)}

\bibitem[{Kegerreis {et~al.}(2019)Kegerreis, Eke, Gonnet, Korycansky, Massey,
  Schaller, \& Teodoro}]{Kegerreis2019}
Kegerreis, J.~A., Eke, V.~R., Gonnet, P., {et~al.} 2019, Monthly Notices of the
  Royal Astronomical Society, 487, 5029, \dodoi{10.1093/mnras/stz1606}

\bibitem[{Kendall \& Melosh(2016)}]{Kendall2016}
Kendall, J.~D., \& Melosh, H.~J. 2016, Earth and Planetary Science Letters,
  448, 24, \dodoi{10.1016/j.epsl.2016.05.012}

\bibitem[{Kraus {et~al.}(2015)Kraus, Root, Lemke, Stewart, Jacobsen, \&
  Mattsson}]{Kraus2015}
Kraus, R.~G., Root, S., Lemke, R.~W., {et~al.} 2015, Nature Geoscience, 8, 269,
  \dodoi{10.1038/ngeo2369}

\bibitem[{{Le Feuvre} \& Wieczorek(2008)}]{LeFeuvre2008}
{Le Feuvre}, M., \& Wieczorek, M.~A. 2008, Icarus, 197, 291,
  \dodoi{10.1016/j.icarus.2008.04.011}

\bibitem[{Liggins {et~al.}(2020)Liggins, Shorttle, \& Rimmer}]{Liggins2020}
Liggins, P., Shorttle, O., \& Rimmer, P.~B. 2020, Earth and Planetary Science
  Letters, 550, 116546, \dodoi{10.1016/j.epsl.2020.116546}

\bibitem[{Marchi {et~al.}(2014)Marchi, Bottke, Elkins-Tanton, Bierhaus,
  Wuennemann, Morbidelli, \& Kring}]{Marchi2014}
Marchi, S., Bottke, W.~F., Elkins-Tanton, L.~T., {et~al.} 2014, Nature, 511,
  578, \dodoi{10.1038/nature13539}

\bibitem[{Marchi {et~al.}(2018)Marchi, Canup, \& Walker}]{Marchi2017}
Marchi, S., Canup, R.~M., \& Walker, R.~J. 2018, Nature Geoscience, 11, 77,
  \dodoi{10.1038/s41561-017-0022-3}

\bibitem[{Marcus(2011)}]{marcus2011role}
Marcus, R.~A. 2011, The role of giant impacts in planet formation and internal
  structure (Harvard University, Ph.D. thesis).
\newblock
  \url{https://search.proquest.com/docview/879049245/abstract/8569841375514A2FPQ/1?accountid=14586}

\bibitem[{Marcus {et~al.}(2009)Marcus, Stewart, Sasselov, \&
  Hernquist}]{marcus2009collisional}
Marcus, R.~A., Stewart, S.~T., Sasselov, D., \& Hernquist, L. 2009, The
  Astrophysical Journal Letters, 700, L118,
  \dodoi{10.1088/0004-637X/700/2/L118}

\bibitem[{Melosh(2007)}]{Melosh2007}
Melosh, H.~J. 2007, Meteoritics and Planetary Science, 42, 2079,
  \dodoi{10.1111/j.1945-5100.2007.tb01009.x}

\bibitem[{Miljkovi{\'{c}} {et~al.}(2013)Miljkovi{\'{c}}, Wieczorek, Collins,
  Laneuville, Neumann, Melosh, Solomon, Phillips, Smith, \&
  Zuber}]{Miljkovic2013}
Miljkovi{\'{c}}, K., Wieczorek, M.~A., Collins, G.~S., {et~al.} 2013, Science,
  342, 724, \dodoi{10.1126/science.1243224}

\bibitem[{Monteux \& Arkani-Hamed(2016)}]{Monteux2016}
Monteux, J., \& Arkani-Hamed, J. 2016, Icarus, 264, 246,
  \dodoi{10.1016/j.icarus.2015.09.040}

\bibitem[{Morbidelli {et~al.}(2018)Morbidelli, Nesvorny, Laurenz, Marchi,
  Rubie, Elkins-Tanton, Wieczorek, \& Jacobson}]{Morbidelli2018}
Morbidelli, A., Nesvorny, D., Laurenz, V., {et~al.} 2018, Icarus, 305, 262,
  \dodoi{10.1016/j.icarus.2017.12.046}

\bibitem[{Morbidelli \& Wood(2015)}]{Morbidelli2015}
Morbidelli, A., \& Wood, B.~J. 2015, in The Early Earth: Accretion and
  Differentiation, Vol. 212 (wiley), 71--82, \dodoi{10.1002/9781118860359.ch4}

\bibitem[{Newsom \& Taylor(1989)}]{Newsom1989}
Newsom, H.~E., \& Taylor, S.~R. 1989, Nature, 338, 360,
  \dodoi{10.1038/338360b0}

\bibitem[{Pierazzo \& Melosh(2000)}]{Pierazzo2000a}
Pierazzo, E., \& Melosh, H.~J. 2000, Icarus, 145, 252,
  \dodoi{10.1006/icar.1999.6332}

\bibitem[{Pierazzo {et~al.}(1997)Pierazzo, Vickery, \& Melosh}]{Pierazzo1997a}
Pierazzo, E., Vickery, A.~M., \& Melosh, H.~J. 1997, Icarus, 127, 408,
  \dodoi{10.1006/icar.1997.5713}

\bibitem[{Poole(1951)}]{poole1951evolution}
Poole, J. H.~J. 1951, Royal Dublin Society

\bibitem[{Rubie {et~al.}(2003)Rubie, Melosh, Reid, Liebske, \&
  Righter}]{Rubie2003}
Rubie, D.~C., Melosh, H.~J., Reid, J.~E., Liebske, C., \& Righter, K. 2003,
  Earth and Planetary Science Letters, 205, 239,
  \dodoi{10.1016/S0012-821X(02)01044-0}

\bibitem[{Rubie {et~al.}(2015)Rubie, Nimmo, \& Melosh}]{Rubie2015}
Rubie, D.~C., Nimmo, F., \& Melosh, H.~J. 2015, {Formation of the Earth's
  Core}, Vol.~9 (Elsevier B.V.), 43--79,
  \dodoi{10.1016/B978-0-444-53802-4.00154-8}

\bibitem[{Ruiz-Bonilla {et~al.}(2021)Ruiz-Bonilla, Eke, Kegerreis, Massey, \&
  Teodoro}]{Ruiz-Bonilla2021}
Ruiz-Bonilla, S., Eke, V.~R., Kegerreis, J.~A., Massey, R.~J., \& Teodoro,
  L.~F. 2021, Monthly Notices of the Royal Astronomical Society, 500, 2861,
  \dodoi{10.1093/mnras/staa3385}

\bibitem[{Sleep(2016)}]{Sleep2016}
Sleep, N.~H. 2016, Geochemistry, Geophysics, Geosystems, 17, 2623,
  \dodoi{10.1002/2016GC006305}

\bibitem[{Sleep {et~al.}(1989)Sleep, Zahnle, Kasting, \& Morowitz}]{Sleep1989}
Sleep, N.~H., Zahnle, K.~J., Kasting, J.~F., \& Morowitz, H.~J. 1989, Nature,
  342, 139, \dodoi{10.1038/342139a0}

\bibitem[{Springel(2005)}]{Springel05}
Springel, V. 2005, Monthly notices of the royal astronomical society, 364,
  1105, \dodoi{10.1111/j.1365-2966.2005.09655.x}

\bibitem[{Stewart(2019)}]{Stewart2019report}
Stewart, S.~T. 2019, {Accessible online:
  https://github.com/ststewart/aneos-forsterite-2019/blob/master/EOS-docs/Stewart-2019-ANEOS\_Modifications.pdf}

\bibitem[{Stewart(2020)}]{stewart2020zenodo-ironalloy}
---. 2020, {Equation of State Model Fe85Si15-ANEOS: Development and
  documentation (Version SLVTv0.2G1)},  Zenodo, \dodoi{10.5281/zenodo.3866550}

\bibitem[{Stewart(2021)}]{Stewart2021}
---. 2021, (in prep.)

\bibitem[{Stewart {et~al.}(2019{\natexlab{a}})Stewart, Davies, Duncan, Lock,
  Root, Townsend, Kraus, Caracas, \& Jacobsen}]{Stewart2019}
Stewart, S.~T., Davies, E.~J., Duncan, M.~S., {et~al.} 2019{\natexlab{a}}.
\newblock \doarXiv{1910.04687}

\bibitem[{Stewart {et~al.}(2019{\natexlab{b}})Stewart, Davies, Duncan, Lock,
  Root, Townsend, Kraus, Caracas, \& Jacobsen}]{stewart2019zenodo-forsterite}
---. 2019{\natexlab{b}}, {Equation of State Model Forsterite-ANEOS- SLVTv1.0G1:
  Documentation and Comparisons}, v1.0.0,  Zenodo,
  \dodoi{10.5281/zenodo.3478631}

\bibitem[{Svetsov(2005)}]{Svetsov2005}
Svetsov, V.~V. 2005, Planetary and Space Science, 53, 1205,
  \dodoi{10.1016/j.pss.2005.04.011}

\bibitem[{Svetsov(2007)}]{Svetsov2007}
---. 2007, Solar System Research, 41, 28, \dodoi{10.1134/S0038094607010030}

\bibitem[{Thompson \& Lauson(1972)}]{Thompson1972}
Thompson, S.~L., \& Lauson, H.~S. 1972, Sandia National Laboratory Report,
  SC-RR-71 0, 113p

\bibitem[{Thompson {et~al.}(2019)Thompson, Lauson, Melosh, Collins, \&
  Stewart}]{Thompson2019}
Thompson, S.~L., Lauson, H.~S., Melosh, H.~J., Collins, G.~S., \& Stewart,
  S.~T. 2019, {M-ANEOS (1.0). Zenodo. https://zenodo.org/record/3525030}, 1.0,
  Zenodo, \dodoi{10.5281/zenodo.3525030}

\bibitem[{Tonks \& Melosh(1993)}]{Tonks1993}
Tonks, W.~B., \& Melosh, H.~J. 1993, Journal of Geophysical Research, 98, 5319,
  \dodoi{10.1029/92JE02726}

\bibitem[{Turcotte \& Schubert(2002)}]{turcotte2002geodynamics}
Turcotte, D.~L., \& Schubert, G. 2002, Geodynamics (Cambridge university press)

\bibitem[{Zahnle {et~al.}(2020)Zahnle, Lupu, Catling, \& Wogan}]{Zahnle2020}
Zahnle, K.~J., Lupu, R., Catling, D.~C., \& Wogan, N. 2020, The Planetary
  Science Journal, 1, 11, \dodoi{10.3847/psj/ab7e2c}

\bibitem[{Zellner(2017)}]{zellner2017cataclysm}
Zellner, N.~E. 2017, Origins of Life and Evolution of Biospheres, 47, 261,
  \dodoi{10.1007/s11084-017-9536-3}

\bibitem[{Zhu {et~al.}(2019)Zhu, Artemieva, Morbidelli, Yin, Becker, \&
  W{\"{u}}nnemann}]{Zhu2019}
Zhu, M.~H., Artemieva, N., Morbidelli, A., {et~al.} 2019, Nature, 571, 226,
  \dodoi{10.1038/s41586-019-1359-0}

\end{thebibliography}
\bibliographystyle{aasjournal}

\appendix

\section{Melt scaling}

\begin{figure}[h!]
\centering
\includegraphics[width=0.8\textwidth,trim={0cm 0.3cm 0 0cm},clip]{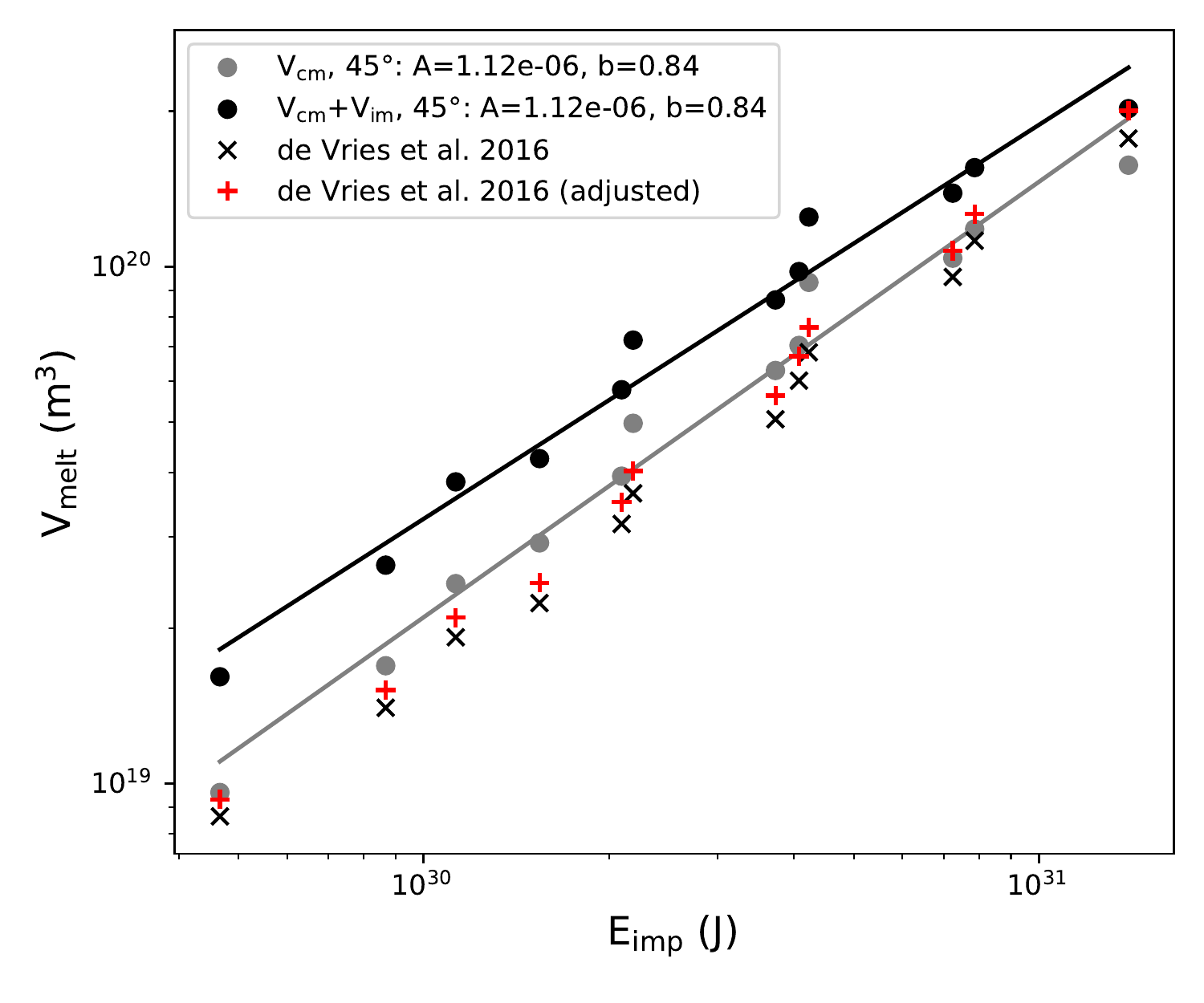}
\caption{Melt volumes vs impact energy for 45$^\circ$ impacts from our GADGET simulation results, compared with an analytic melt scaling from \citet{DeVries2016}. The melt scaling from \citet{DeVries2016} is based on \citet{Abramov2012}. The X's plotted in black use the parameter values from \citet{DeVries2016}, while the adjusted points (red pluses) use slightly different values for a few constant, more similar to our simulation model setup (see text). The grey circles and fit are for the complete melt volume (material with 100\% degree of melting) from our numerical simulations. The black circles and fit are for the total melt volume (complete + incipient melt) from our numerical simulations.}
\label{fig:meltcompare}
\end{figure}

We compare our the impact melt volumes from our simulations with widely the widely used analytical melt scaling from \citet{Abramov2012} (slightly modified by \citet{DeVries2016}) in Figure \ref{fig:meltcompare}. The melt volume as a function of impactor diameter and velocity is given by \citet{Abramov2012} and \citet{DeVries2016} as:

\begin{equation}
V_{\text {melt }}=\frac{\pi}{6} k E_{m}^{-3 \mu / 2} \frac{\rho_{p}}{\rho_{t}} D_{p}^{3} v^{3 \mu} \sin ^{2 \gamma} \theta
\end{equation}

where $V_{\text {melt }}$ is the melt volume, $k$ in an experimentally determined scaling constant, $\mu$ is the energy/momentum velocity scaling exponent, $\rho_p$ and $\rho_t$ are the projectile and target density, respectively, $D_p$ is the projectile diameter, and $\gamma$ is the $\pi$-group scaling parameter ($\gamma = 3\mu / (2+\mu)$). The energy of melting is modified from the form in \citet{Abramov2012} to include a pressure dependence (\citep{DeVries2016}):

\begin{equation}\label{eqn:A2}
    E_{m}=E_{m}^{0}\left(1-\frac{C_{p}\left(T_{s}+\frac{d T}{d z} d_{m}\right)}{C_{p}\left(T_{l 0}+\frac{d T_{l}}{d P} P\right)+L_{m}}\right)
\end{equation}

where $E_m^0$ is the specific energy of melting, $C_p$ is the specific heat at constant pressure, $T_s$ is the surface temperature, $T_{l0}$ is the liquidus temperature at the surface, $d_m$ is the average depth of melting, $P$ is the average pressure of melting, $dT/dz$ is the temperature gradient, $dT_l/dP$ is the change in liquidus temperature with pressure, and $L_m$ is the latent heat of melting. 

The average depth of melting is estimated using a spherical melt volume approximation ($d_m=(3 V_{\text melt}/4\pi)^{1/3}$), which yields \citep{Abramov2012}:

\begin{equation}\label{eqn:A3}
d_m=\frac{1}{2} k^{1/3} E_{m}^{- \mu / 2} \left(\frac{\rho_{p}}{\rho_{t}}\right)^{1/3} D_{p} v^{\mu} \sin ^{2 \gamma/3} \theta
\end{equation}

We solve Equations \ref{eqn:A2} and \ref{eqn:A3} numerically. The pressure $P$ at the depth of melt $d_m$ is given by a simple approximation for a two layer planet $P(r) = \frac{4}{3}\pi \rho_m G b^3 (\rho_c-\rho_m) (\frac{1}{r}-\frac{1}{a})+\frac{2}{3}\pi G \rho_m^2(a^2-r^2)$, where $a$=6371 km is the planet radius, $b$=3486 km is the core radius, $\rho_m$=4000 kg/m$^3$ is the mantle density, $\rho_c$=13200 kg/m$^3$ is the core density, $G$ is the gravitational constant, and $r$ is the radius within the planetary mantle ($b<r<a$) where the pressure is evaluated \citep{turcotte2002geodynamics}. We use an iterative approach to solve for both $d_m$ and $P$. First we assume no pressure dependence in Equation \ref{eqn:A2} to obtain an initial estimate of $d_m$. The initial $d_m$ estimate is used to obtain a value for $P$ that is then used to estimate a new value of $d_m$. This is repeated until the computed value of $d_m$ converges. 

We first use nominal values for constants from \citet{DeVries2016} in order to compare our results with prior analytic melting estimates. We use $k$=0.42, $\mu$=0.56, $\gamma$=0.66, $T_s$=1750 K, $E_m^0$=9.0 MJ/kg, $C_p$=1300 J/kg/K, $L_m$=718 kJ/kg, $T_{l0}$=1950 K, $dT_l/dP|_{P<60 GPa}$=28.3 K/GPa, $dT_l/dP|_{P>60 GPa}$=14.0 K/GPa, and $dT/dz$=0.1 K/km. We assume the projectile and target densities are equal. The melt scaling using these parameters is plotted as black X's in Figure \ref{fig:meltcompare}. We also adjust certain parameters to be more similar to our numerical model and ANEOS melting curve; we use $T_s$=1900 K, $T_{l0}$=2163 K, $dT_l/dP|_{P<60 GPa}$=25 K/GPa, and $dT_l/dP|_{P>60 GPa}$=10 K/GPa. The melt scaling using these adjusted parameters is plotted as the red pluses in Figure \ref{fig:meltcompare}. We use $\theta=45^\circ$ and compare with our 45$^\circ$ impact simulations.

The comparison between our numerical results and the analytical melt scaling for impacts of $45^\circ$ (Figure \ref{fig:meltcompare}) shows that the melt scaling from \citet{Abramov2012} and \citet{DeVries2016} provides a reasonable approximation for the volume of completely melted material (\ie, material with 100\% degree of melting). This is somewhat surprising given that the melt scaling from \citet{Abramov2012} is based on a 2D half space model \citep{Pierazzo2000a} and constructed to fit impacts of diameter $\sim$ 10 km, much smaller than the impacts of up to 3400 km diameter that we model here. However, we do note that the melt scaling relations do not match the total melt produced (complete + incipient melting) in the numerical simulations, as shown in Figure \ref{fig:meltcompare}; the analytical melt scaling estimate appears to underestimate total melt production by a factor of $\sim$ 2 due to the failure to capture the incipient melt (\ie, the mass of melt in only partially melted material for an assumed single phase mantle). Differences between the analytical estimate and our estimated melt may also be due to our simulations using differentiated projectiles in 3D spherical geometry, whereas the melt scaling relations were constructed based on a 2D half-space model for undifferentiated projectiles.

\clearpage
\section{Supplementary Figures}
\setcounter{figure}{0}    
\counterwithin{figure}{section}

\begin{figure}[h!]
\centering
\includegraphics[width=1.0\textwidth,trim={0 0cm 0 0.5cm},clip]{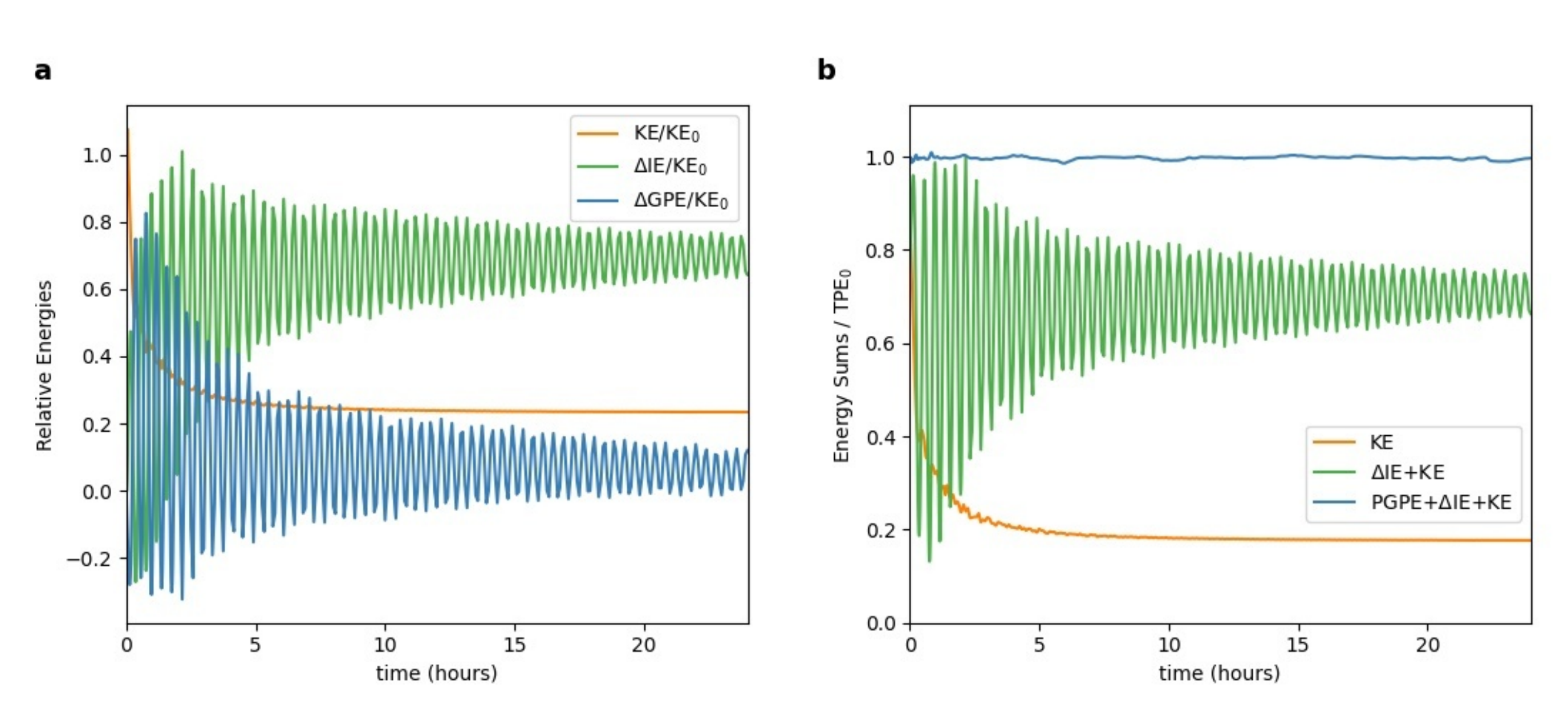}
\caption{Energy conservation plots for an example simulation with $M_i=0.006 M_{Earth}$, $v_{imp}= 1.5 v_{esc}$, and $\theta=45^{\circ}$. (a) Relative energies are shown, displaying the exchange between kinetic energy (KE), the change in internal energy ($\Delta$IE=IE-IE$_0$), and the change in the gravitational potential energy ($\Delta$GPE=GPE-GPE$_0$), all normalized to the initial kinetic energy (KE$_0$). (b) Energy sums are shown, relative to the initial total participating energy (TPE = PGPE + IE + KE). The participating gravitational potential energy is given by PGPE = GPE - min(GPE). }
\label{fig:energyhist}
\end{figure}

\begin{figure}[h!]
\centering
\includegraphics[width=0.9\textwidth,trim={0 0cm 0 0.5cm},clip]{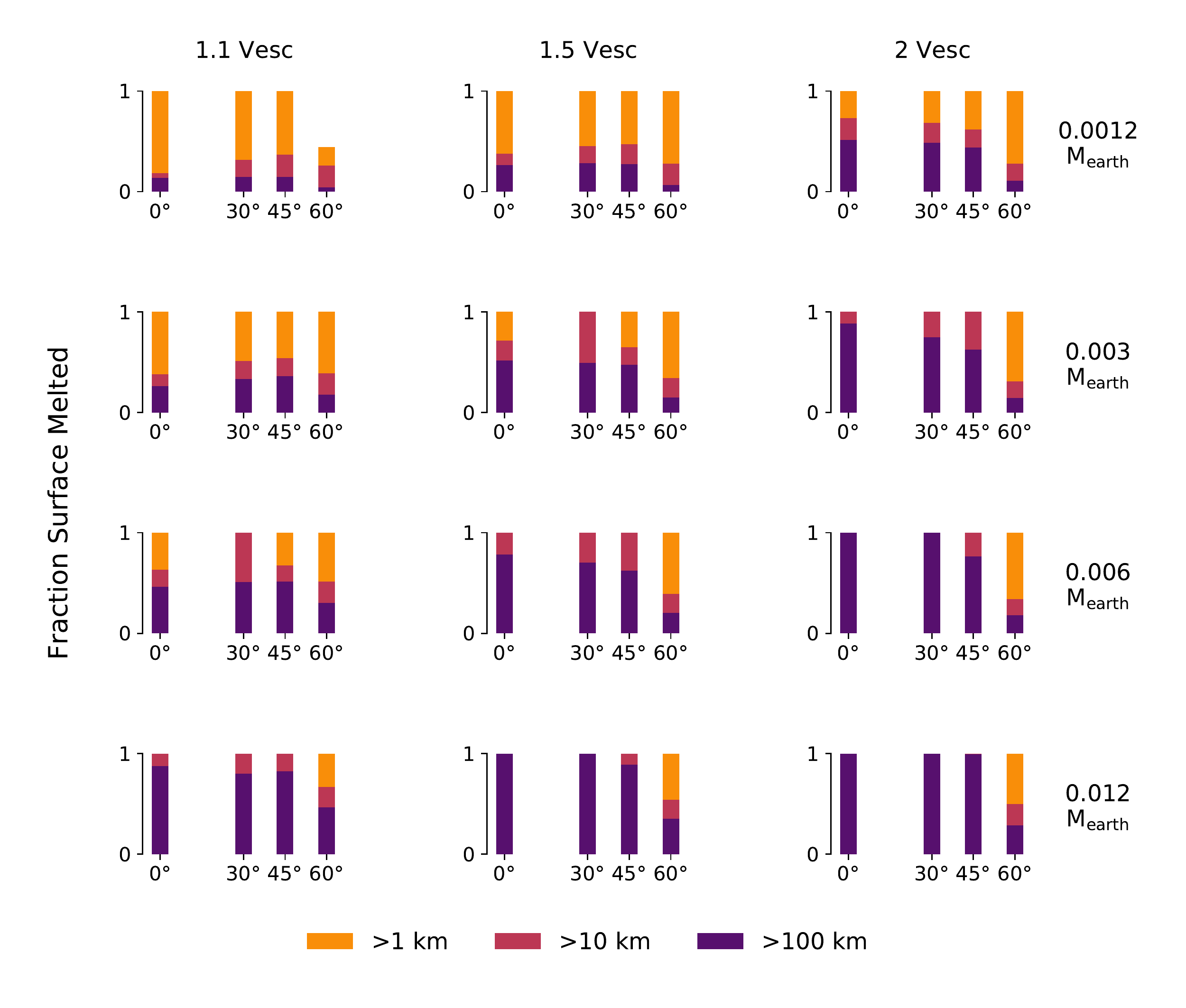}
\caption{Bar charts showing the fraction of surface with $>$ 100, 10, or 1 km of melt after 24 hrs of simulation time. Charts are distributed in columns and rows according to the projectile velocity and mass. The x-axis in each bar chart shows runs of different impact angle (0$^\circ$ = head-on collision). As Figure \ref{fig:stats_surfacemelt} but includes thickness of condensed forsterite atmosphere. }
\label{fig:stats_surfacemelt2}
\end{figure}

\begin{figure}[h!]
\centering
\includegraphics[width=1.0\textwidth,trim={1cm 0.5cm 1cm 0cm},clip]{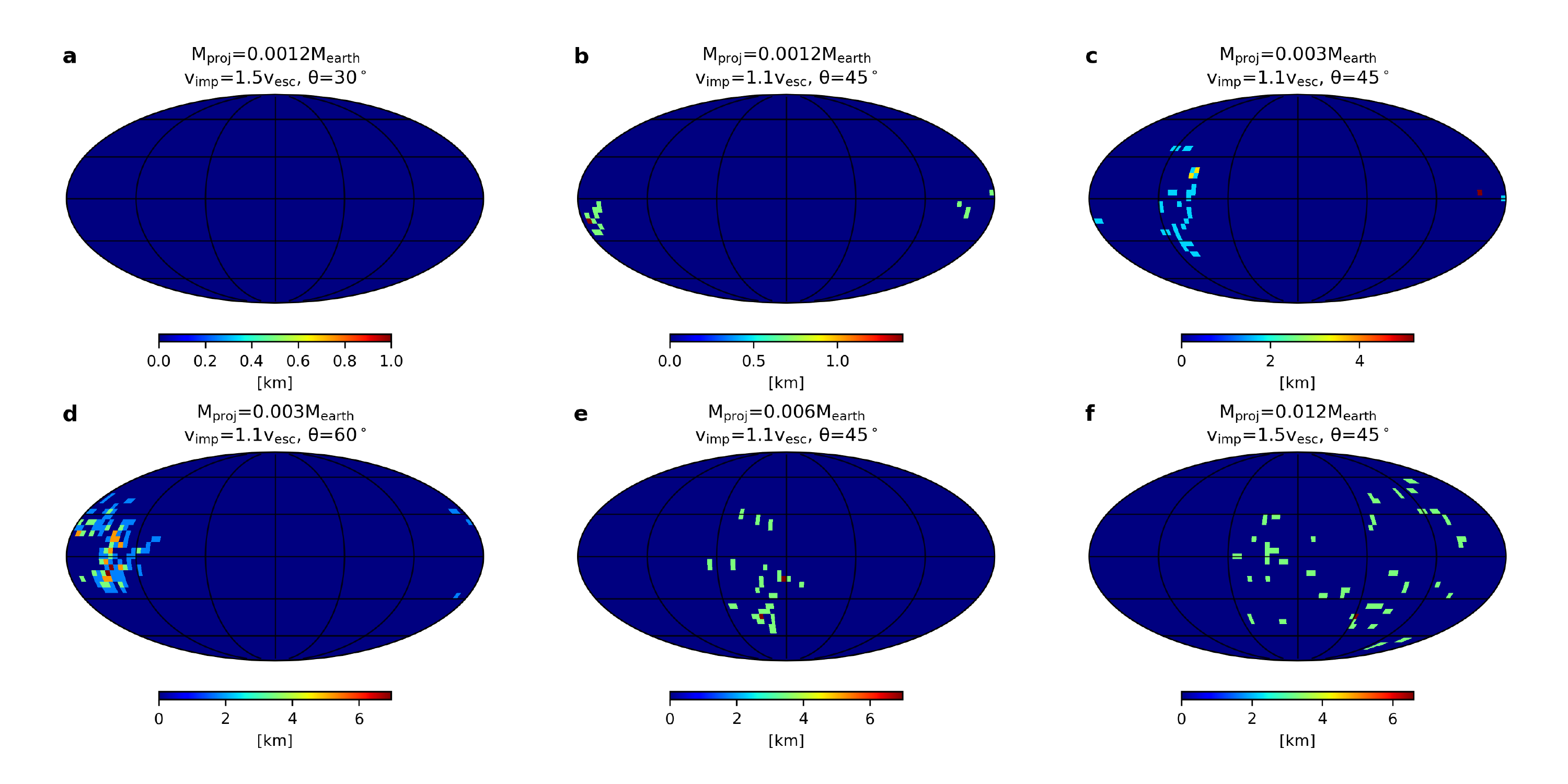}
\caption{Distribution of iron over the surface of the Earth 24 hours post impact. Surface iron is liquid or sold and is plotted as an equivalent layer thickness. The surface layer is defined as the particles with P $<$ 10 GPa and $\rho >$ 1000 kg/m$^3$ within two smoothing lengths of the planetary radius. }
\label{fig:surface_iron}
\end{figure}

\begin{figure}[h!]
\centering
\includegraphics[width=1.0\textwidth,trim={0cm 0cm 0cm 0cm},clip]{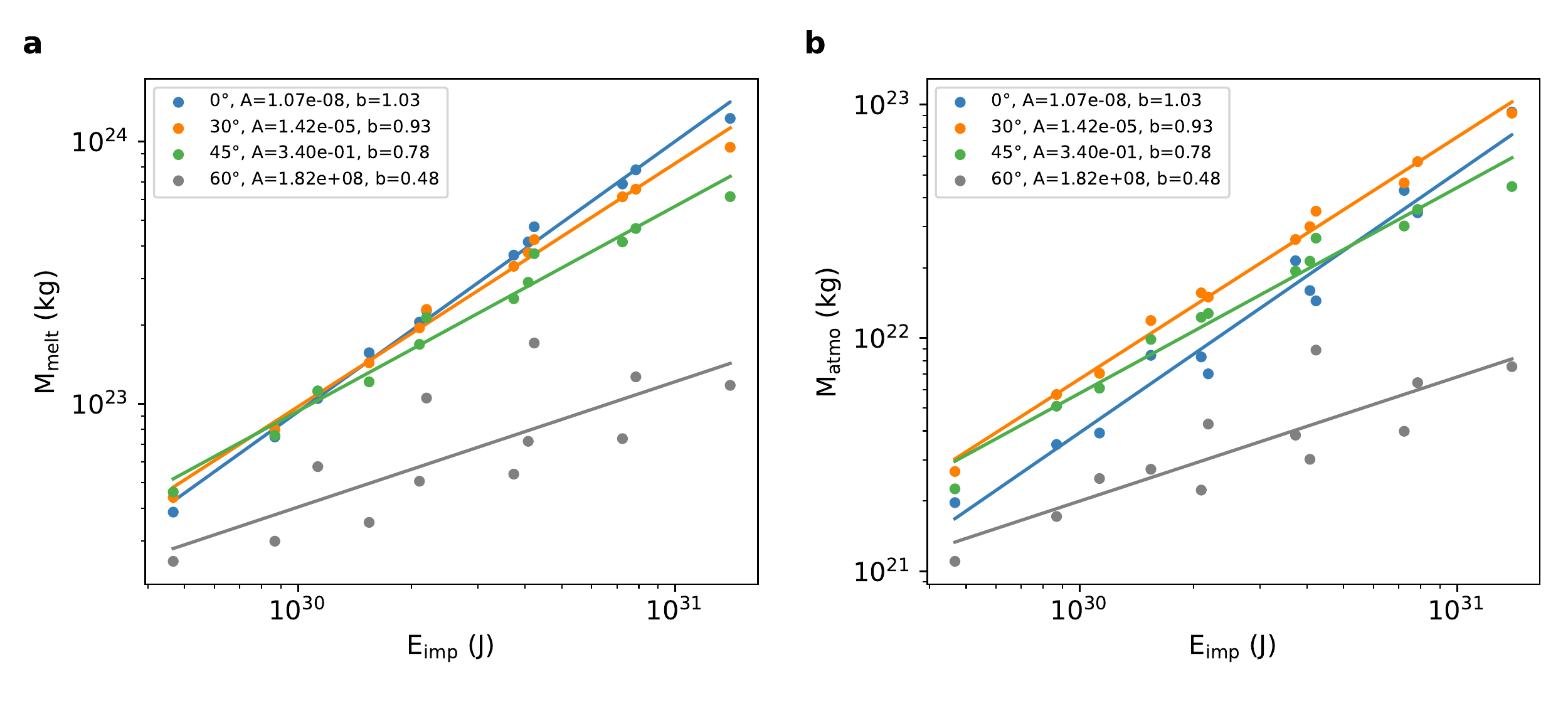}
\caption{(a) Mass of forsterite melt (including supercritical fluid) in the surface and interior. (b) Mass of the forsterite atmosphere (all phases). Results are for after 24 hours of simulation time. See Table \ref{tbl:fors_part}.}
\label{fig:Mmelt}
\end{figure}

\clearpage
\section{Supplementary Data}
\setcounter{table}{0}    
\counterwithin{table}{section}

\begin{table}[!h]
\small
\centering
\caption{Energy partitioning results}
\begin{threeparttable}
\begin{tabular}{ccccccccccc}
\hline
$\frac{M_{proj}}{M_{earth}}$ & $\frac{v_{imp}}{v_{esc}}$ & $\theta$ & $E_{imp}$ [J] & $\frac{\Delta E_{interior}}{E_{imp}}$ & $\frac{\Delta E_{surf}}{E_{imp}}$ & $\frac{\Delta E_{atmo}}{E_{imp}}$ & $\frac{\Delta E_{disk}}{E_{imp}}$ & $\frac{\Delta E_{esc}}{E_{imp}}$ & $\frac{ \Delta IE^{surf}}{E_{imp}}$ & $\frac{ \Delta IE^{atmo}}{E_{imp}}$ \\
\ &\\[-1.0em]
\hline
 0.0012 & 1.1 & 0  & $4.7 \times 10^{29}$ & 0.18 & 0.6  & 0.09 &  0   &  0   & 0.36 & 0.09 \\
 0.0012 & 1.1 & 30 & $4.7 \times 10^{29}$ & 0.23 & 0.58 & 0.11 &  0   & 0.03 & 0.45 & 0.11 \\
 0.0012 & 1.1 & 45 & $4.7 \times 10^{29}$ & 0.14 & 0.55 & 0.08 & 0.01 & 0.15 & 0.47 & 0.09 \\
 0.0012 & 1.1 & 60 & $4.7 \times 10^{29}$ & 0.17 & 0.32 & 0.06 & 0.01 & 0.38 & 0.3  & 0.06 \\
 0.0012 & 1.5 & 0  & $8.7 \times 10^{29}$ & 0.22 & 0.62 & 0.09 &  0   &  0   & 0.3  & 0.09 \\
 0.0012 & 1.5 & 30 & $8.7 \times 10^{29}$ & 0.2  & 0.54 & 0.14 &  0   & 0.05 & 0.34 & 0.14 \\
 0.0012 & 1.5 & 45 & $8.7 \times 10^{29}$ & 0.17 & 0.47 & 0.11 &  0   & 0.21 & 0.34 & 0.12 \\
 0.0012 & 1.5 & 60 & $8.7 \times 10^{29}$ & 0.16 & 0.21 & 0.04 &  0   & 0.56 & 0.18 & 0.04 \\
 0.0012 &  2  & 0  & $1.5 \times 10^{30}$ & 0.16 & 0.7  & 0.12 & 0.01 & 0.03 & 0.27 & 0.12 \\
 0.0012 &  2  & 30 & $1.5 \times 10^{30}$ & 0.17 & 0.56 & 0.19 &  0   & 0.06 & 0.28 & 0.18 \\
 0.0012 &  2  & 45 & $1.5 \times 10^{30}$ & 0.14 & 0.44 & 0.14 & 0.01 & 0.25 & 0.26 & 0.14 \\
 0.0012 &  2  & 60 & $1.5 \times 10^{30}$ & 0.12 & 0.15 & 0.03 &  0   & 0.68 & 0.11 & 0.04 \\
 0.003  & 1.1 & 0  & $1.1 \times 10^{30}$ & 0.16 & 0.69 & 0.07 &  0   &  0   & 0.28 & 0.08 \\
 0.003  & 1.1 & 30 & $1.1 \times 10^{30}$ & 0.19 & 0.61 & 0.12 &  0   & 0.05 & 0.38 & 0.13 \\
 0.003  & 1.1 & 45 & $1.1 \times 10^{30}$ & 0.14 & 0.56 & 0.1  &  0   & 0.19 & 0.38 & 0.11 \\
 0.003  & 1.1 & 60 & $1.1 \times 10^{30}$ & 0.19 & 0.3  & 0.05 & 0.02 & 0.42 & 0.23 & 0.06 \\
 0.003  & 1.5 & 0  & $2.1 \times 10^{30}$ & 0.18 & 0.7  & 0.1  &  0   &  0   & 0.25 & 0.1  \\
 0.003  & 1.5 & 30 & $2.1 \times 10^{30}$ & 0.2  & 0.53 & 0.17 &  0   & 0.06 & 0.27 & 0.17 \\
 0.003  & 1.5 & 45 & $2.1 \times 10^{30}$ & 0.17 & 0.42 & 0.13 & 0.01 & 0.26 & 0.26 & 0.13 \\
 0.003  & 1.5 & 60 & $2.1 \times 10^{30}$ & 0.17 & 0.16 & 0.02 &  0   & 0.63 & 0.11 & 0.03 \\
 0.003  &  2  & 0  & $3.7 \times 10^{30}$ & 0.16 & 0.68 & 0.15 &  0   &  0   & 0.22 & 0.14 \\
 0.003  &  2  & 30 & $3.7 \times 10^{30}$ & 0.18 & 0.53 & 0.19 &  0   & 0.09 & 0.22 & 0.18 \\
 0.003  &  2  & 45 & $3.7 \times 10^{30}$ & 0.16 & 0.37 & 0.13 & 0.01 & 0.33 & 0.19 & 0.13 \\
 0.003  &  2  & 60 & $3.7 \times 10^{30}$ & 0.11 & 0.1  & 0.02 &  0   & 0.76 & 0.06 & 0.02 \\
 0.006  & 1.1 & 0  & $2.2 \times 10^{30}$ & 0.26 & 0.62 & 0.07 &  0   &  0   & 0.23 & 0.07 \\
 0.006  & 1.1 & 30 & $2.2 \times 10^{30}$ & 0.25 & 0.52 & 0.15 &  0   & 0.06 & 0.3  & 0.15 \\
 0.006  & 1.1 & 45 & $2.2 \times 10^{30}$ & 0.21 & 0.45 & 0.12 & 0.01 & 0.21 & 0.3  & 0.13 \\
 0.006  & 1.1 & 60 & $2.2 \times 10^{30}$ & 0.25 & 0.22 & 0.05 & 0.01 & 0.46 & 0.16 & 0.05 \\
 0.006  & 1.5 & 0  & $4.1 \times 10^{30}$ & 0.3  & 0.59 & 0.11 &  0   &  0   & 0.21 & 0.11 \\
 0.006  & 1.5 & 30 & $4.1 \times 10^{30}$ & 0.27 & 0.44 & 0.19 &  0   & 0.09 & 0.22 & 0.18 \\
 0.006  & 1.5 & 45 & $4.1 \times 10^{30}$ & 0.23 & 0.32 & 0.13 & 0.01 & 0.31 & 0.19 & 0.13 \\
 0.006  & 1.5 & 60 & $4.1 \times 10^{30}$ & 0.2  & 0.1  & 0.01 &  0   & 0.68 & 0.07 & 0.02 \\
 0.006  &  2  & 0  & $7.2 \times 10^{30}$ & 0.31 & 0.52 & 0.18 &  0   & 0.01 & 0.18 & 0.16 \\
 0.006  &  2  & 30 & $7.2 \times 10^{30}$ & 0.26 & 0.42 & 0.18 & 0.02 & 0.12 & 0.19 & 0.17 \\
 0.006  &  2  & 45 & $7.2 \times 10^{30}$ & 0.21 & 0.27 & 0.12 & 0.01 & 0.41 & 0.14 & 0.11 \\
 0.006  &  2  & 60 & $7.2 \times 10^{30}$ & 0.12 & 0.06 & 0.01 &  0   & 0.8  & 0.03 & 0.01 \\
 0.012  & 1.1 & 0  & $4.2 \times 10^{30}$ & 0.28 & 0.61 & 0.09 &  0   & 0.01 & 0.21 & 0.09 \\
 0.012  & 1.1 & 30 & $4.2 \times 10^{30}$ & 0.24 & 0.45 & 0.22 & 0.01 & 0.08 & 0.26 & 0.21 \\
 0.012  & 1.1 & 45 & $4.2 \times 10^{30}$ & 0.2  & 0.39 & 0.15 & 0.01 & 0.25 & 0.27 & 0.15 \\
 0.012  & 1.1 & 60 & $4.2 \times 10^{30}$ & 0.27 & 0.18 & 0.05 & 0.02 & 0.48 & 0.13 & 0.06 \\
 0.012  & 1.5 & 0  & $7.9 \times 10^{30}$ & 0.36 & 0.48 & 0.14 &  0   & 0.02 & 0.18 & 0.13 \\
 0.012  & 1.5 & 30 & $7.9 \times 10^{30}$ & 0.28 & 0.38 & 0.22 & 0.01 & 0.13 & 0.2  & 0.2  \\
 0.012  & 1.5 & 45 & $7.9 \times 10^{30}$ & 0.23 & 0.25 & 0.13 & 0.01 & 0.38 & 0.16 & 0.12 \\
 0.012  & 1.5 & 60 & $7.9 \times 10^{30}$ & 0.19 & 0.09 & 0.02 &  0   & 0.71 & 0.05 & 0.02 \\
 0.012  &  2  & 0  & $1.4 \times 10^{31}$ & 0.38 & 0.37 & 0.23 & 0.01 & 0.03 & 0.14 & 0.21 \\
 0.012  &  2  & 30 & $1.4 \times 10^{31}$ & 0.3  & 0.3  & 0.21 & 0.02 & 0.18 & 0.16 & 0.19 \\
 0.012  &  2  & 45 & $1.4 \times 10^{31}$ & 0.2  & 0.2  & 0.1  & 0.01 & 0.5  & 0.11 & 0.09 \\
 0.012  &  2  & 60 & $1.4 \times 10^{31}$ & 0.12 & 0.05 & 0.01 &  0   & 0.83 & 0.03 & 0.01 \\
\hline
\end{tabular}

\begin{tablenotes}\footnotesize
\item[] \textit{Notes:}
\item[1.] $\Delta E_{interior}$, $\Delta E_{atmo}$, $\Delta E_{surf}$, $\Delta E_{disk}$ and $\Delta E_{esc}$ are change in energy for particles composing the interior, atmosphere, surface, disk, and unbound material, after 24 hours of simulation time, and are normalized to the impact energy $E_{imp}$. $\Delta E = (IE_{final}-IE_{initial})+(GPE_{final}-GPE_{initial})+KE_{final}$, where $IE$, $GPE$ and $KE$ are the internal energy, gravitational potential energy, and kinetic energy, respectively.  
\item[2.] $\Delta IE^{surf}$ and $\Delta IE^{atmo}$ are the change in the internal energy of surface and atmospheric particles, after 24 hours of simulation time. 
\end{tablenotes}
\end{threeparttable}
\label{tbl:energy_part}
\end{table}

\begin{table}[!h]
\small
\centering
\caption{Forsterite melting and partitioning}
\begin{threeparttable}
\begin{tabular}{cccccccccc}
\toprule
$\frac{M_{proj}}{M_{earth}}$ & $\frac{v_{imp}}{v_{esc}}$ & $\theta$ & $E_{imp}$ [J] & $M_{melt}$ [kg] & $M_{scf}$ [kg] & $M_{scf}^{atmo}$ [kg] & $M_{vapor}$ [kg] & $M_{atmo}$ [kg] & $M_{disk}$ [kg]\\
\\[-1.0em]
\hline
 0.0012 & 1.1 & 0  & $4.7 \times 10^{29}$ & $3.8 \times 10^{22}$ & $2.0 \times 10^{21}$ & $1.4 \times 10^{21}$ & $5.8 \times 10^{20}$ & $2.0 \times 10^{21}$ & $1.1 \times 10^{19}$ \\
 0.0012 & 1.1 & 30 & $4.7 \times 10^{29}$ & $4.3 \times 10^{22}$ & $3.5 \times 10^{21}$ & $2.2 \times 10^{21}$ & $4.7 \times 10^{20}$ & $2.7 \times 10^{21}$ & $5.6 \times 10^{19}$ \\
 0.0012 & 1.1 & 45 & $4.7 \times 10^{29}$ & $4.5 \times 10^{22}$ & $3.3 \times 10^{21}$ & $2.0 \times 10^{21}$ & $2.4 \times 10^{20}$ & $2.3 \times 10^{21}$ & $1.0 \times 10^{20}$ \\
 0.0012 & 1.1 & 60 & $4.7 \times 10^{29}$ & $2.5 \times 10^{22}$ & $1.3 \times 10^{21}$ & $9.6 \times 10^{20}$ & $1.4 \times 10^{20}$ & $1.1 \times 10^{21}$ & $1.5 \times 10^{20}$ \\
 0.0012 & 1.5 & 0  & $8.7 \times 10^{29}$ & $7.3 \times 10^{22}$ & $3.6 \times 10^{21}$ & $1.9 \times 10^{21}$ & $1.6 \times 10^{21}$ & $3.5 \times 10^{21}$ & $8.7 \times 10^{18}$ \\
 0.0012 & 1.5 & 30 & $8.7 \times 10^{29}$ & $7.8 \times 10^{22}$ & $5.7 \times 10^{21}$ & $3.7 \times 10^{21}$ & $2.1 \times 10^{21}$ & $5.7 \times 10^{21}$ & $7.2 \times 10^{19}$ \\
 0.0012 & 1.5 & 45 & $8.7 \times 10^{29}$ & $7.4 \times 10^{22}$ & $5.5 \times 10^{21}$ & $3.9 \times 10^{21}$ & $1.2 \times 10^{21}$ & $5.1 \times 10^{21}$ & $1.3 \times 10^{20}$ \\
 0.0012 & 1.5 & 60 & $8.7 \times 10^{29}$ & $2.9 \times 10^{22}$ & $2.2 \times 10^{21}$ & $1.5 \times 10^{21}$ & $2.5 \times 10^{20}$ & $1.7 \times 10^{21}$ & $9.7 \times 10^{19}$ \\
 0.0012 &  2  & 0  & $1.5 \times 10^{30}$ & $1.5 \times 10^{23}$ & $9.7 \times 10^{21}$ & $4.9 \times 10^{21}$ & $3.5 \times 10^{21}$ & $8.4 \times 10^{21}$ & $2.1 \times 10^{20}$ \\
 0.0012 &  2  & 30 & $1.5 \times 10^{30}$ & $1.4 \times 10^{23}$ & $9.9 \times 10^{21}$ & $6.2 \times 10^{21}$ & $5.6 \times 10^{21}$ & $1.2 \times 10^{22}$ & $1.1 \times 10^{20}$ \\
 0.0012 &  2  & 45 & $1.5 \times 10^{30}$ & $1.2 \times 10^{23}$ & $8.9 \times 10^{21}$ & $6.4 \times 10^{21}$ & $3.5 \times 10^{21}$ & $9.9 \times 10^{21}$ & $2.7 \times 10^{20}$ \\
 0.0012 &  2  & 60 & $1.5 \times 10^{30}$ & $3.5 \times 10^{22}$ & $2.7 \times 10^{21}$ & $2.2 \times 10^{21}$ & $5.8 \times 10^{20}$ & $2.7 \times 10^{21}$ & $8.5 \times 10^{19}$ \\
 0.003  & 1.1 & 0  & $1.1 \times 10^{30}$ & $1.0 \times 10^{23}$ & $5.2 \times 10^{21}$ & $3.4 \times 10^{21}$ & $4.9 \times 10^{20}$ & $3.9 \times 10^{21}$ & $1.9 \times 10^{19}$ \\
 0.003  & 1.1 & 30 & $1.1 \times 10^{30}$ & $1.1 \times 10^{23}$ & $1.1 \times 10^{22}$ & $6.8 \times 10^{21}$ & $2.7 \times 10^{20}$ & $7.1 \times 10^{21}$ & $1.2 \times 10^{20}$ \\
 0.003  & 1.1 & 45 & $1.1 \times 10^{30}$ & $1.1 \times 10^{23}$ & $9.3 \times 10^{21}$ & $6.0 \times 10^{21}$ & $5.7 \times 10^{19}$ & $6.1 \times 10^{21}$ & $1.6 \times 10^{20}$ \\
 0.003  & 1.1 & 60 & $1.1 \times 10^{30}$ & $5.7 \times 10^{22}$ & $3.4 \times 10^{21}$ & $2.4 \times 10^{21}$ & $6.1 \times 10^{19}$ & $2.5 \times 10^{21}$ & $5.1 \times 10^{20}$ \\
 0.003  & 1.5 & 0  & $2.1 \times 10^{30}$ & $2.0 \times 10^{23}$ & $1.2 \times 10^{22}$ & $5.5 \times 10^{21}$ & $2.8 \times 10^{21}$ & $8.3 \times 10^{21}$ & $2.0 \times 10^{19}$ \\
 0.003  & 1.5 & 30 & $2.1 \times 10^{30}$ & $1.9 \times 10^{23}$ & $1.8 \times 10^{22}$ & $1.3 \times 10^{22}$ & $2.7 \times 10^{21}$ & $1.6 \times 10^{22}$ & $2.0 \times 10^{20}$ \\
 0.003  & 1.5 & 45 & $2.1 \times 10^{30}$ & $1.6 \times 10^{23}$ & $1.5 \times 10^{22}$ & $1.1 \times 10^{22}$ & $8.4 \times 10^{20}$ & $1.2 \times 10^{22}$ & $3.7 \times 10^{20}$ \\
 0.003  & 1.5 & 60 & $2.1 \times 10^{30}$ & $4.9 \times 10^{22}$ & $4.2 \times 10^{21}$ & $2.1 \times 10^{21}$ & $8.1 \times 10^{19}$ & $2.2 \times 10^{21}$ & $2.0 \times 10^{20}$ \\
 0.003  &  2  & 0  & $3.7 \times 10^{30}$ & $3.5 \times 10^{23}$ & $3.2 \times 10^{22}$ & $1.5 \times 10^{22}$ & $6.1 \times 10^{21}$ & $2.1 \times 10^{22}$ & $1.4 \times 10^{20}$ \\
 0.003  &  2  & 30 & $3.7 \times 10^{30}$ & $3.2 \times 10^{23}$ & $2.9 \times 10^{22}$ & $1.8 \times 10^{22}$ & $8.0 \times 10^{21}$ & $2.6 \times 10^{22}$ & $4.7 \times 10^{20}$ \\
 0.003  &  2  & 45 & $3.7 \times 10^{30}$ & $2.5 \times 10^{23}$ & $2.2 \times 10^{22}$ & $1.6 \times 10^{22}$ & $3.5 \times 10^{21}$ & $1.9 \times 10^{22}$ & $5.6 \times 10^{20}$ \\
 0.003  &  2  & 60 & $3.7 \times 10^{30}$ & $5.2 \times 10^{22}$ & $5.9 \times 10^{21}$ & $3.7 \times 10^{21}$ & $1.1 \times 10^{20}$ & $3.8 \times 10^{21}$ & $1.9 \times 10^{20}$ \\
 0.006  & 1.1 & 0  & $2.2 \times 10^{30}$ & $2.2 \times 10^{23}$ & $1.2 \times 10^{22}$ & $6.8 \times 10^{21}$ & $2.1 \times 10^{20}$ & $7.0 \times 10^{21}$ & $1.4 \times 10^{20}$ \\
 0.006  & 1.1 & 30 & $2.2 \times 10^{30}$ & $2.2 \times 10^{23}$ & $2.4 \times 10^{22}$ & $1.4 \times 10^{22}$ & $5.0 \times 10^{20}$ & $1.5 \times 10^{22}$ & $2.4 \times 10^{20}$ \\
 0.006  & 1.1 & 45 & $2.2 \times 10^{30}$ & $2.1 \times 10^{23}$ & $1.9 \times 10^{22}$ & $1.3 \times 10^{22}$ & $6.9 \times 10^{19}$ & $1.3 \times 10^{22}$ & $3.7 \times 10^{20}$ \\
 0.006  & 1.1 & 60 & $2.2 \times 10^{30}$ & $1.0 \times 10^{23}$ & $6.4 \times 10^{21}$ & $4.2 \times 10^{21}$ & $5.9 \times 10^{19}$ & $4.3 \times 10^{21}$ & $8.6 \times 10^{20}$ \\
 0.006  & 1.5 & 0  & $4.1 \times 10^{30}$ & $4.0 \times 10^{23}$ & $3.0 \times 10^{22}$ & $1.1 \times 10^{22}$ & $4.8 \times 10^{21}$ & $1.6 \times 10^{22}$ & $9.3 \times 10^{19}$ \\
 0.006  & 1.5 & 30 & $4.1 \times 10^{30}$ & $3.6 \times 10^{23}$ & $3.8 \times 10^{22}$ & $2.5 \times 10^{22}$ & $4.7 \times 10^{21}$ & $3.0 \times 10^{22}$ & $5.1 \times 10^{20}$ \\
 0.006  & 1.5 & 45 & $4.1 \times 10^{30}$ & $2.8 \times 10^{23}$ & $3.0 \times 10^{22}$ & $2.1 \times 10^{22}$ & $7.9 \times 10^{20}$ & $2.1 \times 10^{22}$ & $7.3 \times 10^{20}$ \\
 0.006  & 1.5 & 60 & $4.1 \times 10^{30}$ & $6.9 \times 10^{22}$ & $5.6 \times 10^{21}$ & $3.0 \times 10^{21}$ & $6.2 \times 10^{19}$ & $3.0 \times 10^{21}$ & $3.1 \times 10^{20}$ \\
 0.006  &  2  & 0  & $7.2 \times 10^{30}$ & $6.4 \times 10^{23}$ & $7.4 \times 10^{22}$ & $3.0 \times 10^{22}$ & $1.3 \times 10^{22}$ & $4.3 \times 10^{22}$ & $4.2 \times 10^{20}$ \\
 0.006  &  2  & 30 & $7.2 \times 10^{30}$ & $5.8 \times 10^{23}$ & $6.9 \times 10^{22}$ & $3.7 \times 10^{22}$ & $9.5 \times 10^{21}$ & $4.6 \times 10^{22}$ & $3.4 \times 10^{21}$ \\
 0.006  &  2  & 45 & $7.2 \times 10^{30}$ & $4.0 \times 10^{23}$ & $4.1 \times 10^{22}$ & $2.6 \times 10^{22}$ & $3.8 \times 10^{21}$ & $3.0 \times 10^{22}$ & $1.1 \times 10^{21}$ \\
 0.006  &  2  & 60 & $7.2 \times 10^{30}$ & $7.0 \times 10^{22}$ & $7.7 \times 10^{21}$ & $3.9 \times 10^{21}$ & $3.8 \times 10^{19}$ & $4.0 \times 10^{21}$ & $2.4 \times 10^{20}$ \\
 0.012  & 1.1 & 0  & $4.2 \times 10^{30}$ & $4.6 \times 10^{23}$ & $2.7 \times 10^{22}$ & $1.2 \times 10^{22}$ & $2.5 \times 10^{21}$ & $1.4 \times 10^{22}$ & $7.8 \times 10^{19}$ \\
 0.012  & 1.1 & 30 & $4.2 \times 10^{30}$ & $4.1 \times 10^{23}$ & $4.8 \times 10^{22}$ & $3.0 \times 10^{22}$ & $4.8 \times 10^{21}$ & $3.5 \times 10^{22}$ & $7.5 \times 10^{20}$ \\
 0.012  & 1.1 & 45 & $4.2 \times 10^{30}$ & $3.6 \times 10^{23}$ & $4.4 \times 10^{22}$ & $2.6 \times 10^{22}$ & $1.1 \times 10^{21}$ & $2.7 \times 10^{22}$ & $1.2 \times 10^{21}$ \\
 0.012  & 1.1 & 60 & $4.2 \times 10^{30}$ & $1.7 \times 10^{23}$ & $1.3 \times 10^{22}$ & $8.7 \times 10^{21}$ & $1.4 \times 10^{20}$ & $8.9 \times 10^{21}$ & $1.5 \times 10^{21}$ \\
 0.012  & 1.5 & 0  & $7.9 \times 10^{30}$ & $7.3 \times 10^{23}$ & $7.1 \times 10^{22}$ & $2.3 \times 10^{22}$ & $1.2 \times 10^{22}$ & $3.4 \times 10^{22}$ & $7.4 \times 10^{20}$ \\
 0.012  & 1.5 & 30 & $7.9 \times 10^{30}$ & $6.2 \times 10^{23}$ & $8.1 \times 10^{22}$ & $4.4 \times 10^{22}$ & $1.3 \times 10^{22}$ & $5.7 \times 10^{22}$ & $1.8 \times 10^{21}$ \\
 0.012  & 1.5 & 45 & $7.9 \times 10^{30}$ & $4.5 \times 10^{23}$ & $5.0 \times 10^{22}$ & $3.2 \times 10^{22}$ & $3.7 \times 10^{21}$ & $3.6 \times 10^{22}$ & $1.7 \times 10^{21}$ \\
 0.012  & 1.5 & 60 & $7.9 \times 10^{30}$ & $1.2 \times 10^{23}$ & $1.2 \times 10^{22}$ & $6.3 \times 10^{21}$ & $9.2 \times 10^{19}$ & $6.4 \times 10^{21}$ & $6.3 \times 10^{20}$ \\
 0.012  &  2  & 0  & $1.4 \times 10^{31}$ & $1.1 \times 10^{24}$ & $1.6 \times 10^{23}$ & $6.5 \times 10^{22}$ & $2.8 \times 10^{22}$ & $9.3 \times 10^{22}$ & $2.1 \times 10^{21}$ \\
 0.012  &  2  & 30 & $1.4 \times 10^{31}$ & $8.7 \times 10^{23}$ & $1.6 \times 10^{23}$ & $7.2 \times 10^{22}$ & $2.0 \times 10^{22}$ & $9.2 \times 10^{22}$ & $5.6 \times 10^{21}$ \\
 0.012  &  2  & 45 & $1.4 \times 10^{31}$ & $5.9 \times 10^{23}$ & $6.6 \times 10^{22}$ & $3.7 \times 10^{22}$ & $7.7 \times 10^{21}$ & $4.5 \times 10^{22}$ & $2.0 \times 10^{21}$ \\
 0.012  &  2  & 60 & $1.4 \times 10^{31}$ & $1.1 \times 10^{23}$ & $1.3 \times 10^{22}$ & $7.5 \times 10^{21}$ & $5.1 \times 10^{19}$ & $7.5 \times 10^{21}$ & $3.0 \times 10^{20}$ \\
\hline
\end{tabular}
\begin{tablenotes}\footnotesize
\item[] \textit{Notes:}
\item[1.] $M_{melt}$, $M_{scf}$, and $M_{vapor}$ give the total liquid, super-critical fluid, and vapor mass, respectively, for forsterite particles within the interior, surface, and atmosphere (all particles except disk and escaping particles).
\item[2.] $M_{scf}^{atmo}$ gives the mass of super-critical fluid in the atmosphere. $M_{scf}$-$M_{scf}^{atmo}$ therefore gives the mass of super-critical fluid in the surface and interior. 
\item[3.] $M_{atmo}$ and $M_{disk}$ gives the total mass of forsterite (of all phases) in the atmosphere and disk. 
\end{tablenotes}
\end{threeparttable}
\label{tbl:fors_part}
\end{table}

\begin{table}[!h]
\small
\centering
\caption{Iron partitioning}
\begin{threeparttable}
\begin{tabular}{cccccccccc}
\toprule
$\frac{M_{proj}}{M_{earth}}$ & $\frac{v_{imp}}{v_{esc}}$ & $\theta$ & $E_{imp}$ [J] & $M_{proj}^{iron}$ [kg]& $X^{Fe}_{interior}$ & $X^{Fe}_{surf}$ & $X^{Fe}_{atmo}$ & $X^{Fe}_{disk}$ & $X^{Fe}_{ejec}$\\
\ &\\[-1.0em]
\hline
 0.0012 & 1.1 & 0  & $4.7 \times 10^{29}$ & 2.3825e+21 &   1   &   0   &   0   &   0   &   0   \\
 0.0012 & 1.1 & 30 & $4.7 \times 10^{29}$ & 2.3825e+21 &   1   &   0   &   0   &   0   &   0   \\
 0.0012 & 1.1 & 45 & $4.7 \times 10^{29}$ & 2.3825e+21 & 0.99  & 0.007 & 0.003 &   0   &   0   \\
 0.0012 & 1.1 & 60 & $4.7 \times 10^{29}$ & 2.3825e+21 & 0.375 & 0.061 & 0.196 & 0.038 & 0.33  \\
 0.0012 & 1.5 & 0  & $8.7 \times 10^{29}$ & 2.3825e+21 &   1   &   0   &   0   &   0   &   0   \\
 0.0012 & 1.5 & 30 & $8.7 \times 10^{29}$ & 2.3825e+21 &   1   &   0   &   0   &   0   &   0   \\
 0.0012 & 1.5 & 45 & $8.7 \times 10^{29}$ & 2.3825e+21 & 0.879 & 0.03  & 0.091 & 0.001 &   0   \\
 0.0012 & 1.5 & 60 & $8.7 \times 10^{29}$ & 2.3825e+21 & 0.156 & 0.062 & 0.177 & 0.018 & 0.588 \\
 0.0012 &  2  & 0  & $1.5 \times 10^{30}$ & 2.3825e+21 & 0.948 & 0.047 & 0.005 &   0   &   0   \\
 0.0012 &  2  & 30 & $1.5 \times 10^{30}$ & 2.3825e+21 & 0.851 & 0.14  & 0.009 &   0   &   0   \\
 0.0012 &  2  & 45 & $1.5 \times 10^{30}$ & 2.3825e+21 & 0.689 & 0.101 & 0.195 & 0.004 & 0.011 \\
 0.0012 &  2  & 60 & $1.5 \times 10^{30}$ & 2.3825e+21 & 0.037 & 0.026 & 0.088 & 0.014 & 0.834 \\
 0.003  & 1.1 & 0  & $1.1 \times 10^{30}$ & 5.9626e+21 &   1   &   0   &   0   &   0   &   0   \\
 0.003  & 1.1 & 30 & $1.1 \times 10^{30}$ & 5.9626e+21 &   1   &   0   &   0   &   0   &   0   \\
 0.003  & 1.1 & 45 & $1.1 \times 10^{30}$ & 5.9626e+21 & 0.96  & 0.018 & 0.022 &   0   &   0   \\
 0.003  & 1.1 & 60 & $1.1 \times 10^{30}$ & 5.9626e+21 & 0.29  & 0.065 & 0.191 & 0.041 & 0.413 \\
 0.003  & 1.5 & 0  & $2.1 \times 10^{30}$ & 5.9626e+21 &   1   &   0   &   0   &   0   &   0   \\
 0.003  & 1.5 & 30 & $2.1 \times 10^{30}$ & 5.9626e+21 &   1   &   0   &   0   &   0   &   0   \\
 0.003  & 1.5 & 45 & $2.1 \times 10^{30}$ & 5.9626e+21 & 0.786 & 0.024 & 0.174 & 0.004 & 0.012 \\
 0.003  & 1.5 & 60 & $2.1 \times 10^{30}$ & 5.9626e+21 & 0.055 & 0.022 & 0.124 & 0.016 & 0.783 \\
 0.003  &  2  & 0  & $3.7 \times 10^{30}$ & 5.9626e+21 & 0.987 & 0.012 &   0   &   0   &   0   \\
 0.003  &  2  & 30 & $3.7 \times 10^{30}$ & 5.9626e+21 & 0.907 & 0.091 & 0.002 &   0   &   0   \\
 0.003  &  2  & 45 & $3.7 \times 10^{30}$ & 5.9626e+21 & 0.543 & 0.053 & 0.287 & 0.004 & 0.113 \\
 0.003  &  2  & 60 & $3.7 \times 10^{30}$ & 5.9626e+21 & 0.002 & 0.007 & 0.03  & 0.011 & 0.95  \\
 0.006  & 1.1 & 0  & $2.2 \times 10^{30}$ & 1.1897e+22 &   1   &   0   &   0   &   0   &   0   \\
 0.006  & 1.1 & 30 & $2.2 \times 10^{30}$ & 1.1897e+22 &   1   &   0   &   0   &   0   &   0   \\
 0.006  & 1.1 & 45 & $2.2 \times 10^{30}$ & 1.1897e+22 & 0.919 & 0.015 & 0.066 &   0   &   0   \\
 0.006  & 1.1 & 60 & $2.2 \times 10^{30}$ & 1.1897e+22 & 0.24  & 0.032 & 0.164 & 0.024 & 0.54  \\
 0.006  & 1.5 & 0  & $4.1 \times 10^{30}$ & 1.1897e+22 &   1   &   0   &   0   &   0   &   0   \\
 0.006  & 1.5 & 30 & $4.1 \times 10^{30}$ & 1.1897e+22 & 0.999 & 0.001 &   0   &   0   &   0   \\
 0.006  & 1.5 & 45 & $4.1 \times 10^{30}$ & 1.1897e+22 & 0.652 & 0.032 & 0.24  & 0.018 & 0.059 \\
 0.006  & 1.5 & 60 & $4.1 \times 10^{30}$ & 1.1897e+22 & 0.013 & 0.01  & 0.053 & 0.012 & 0.912 \\
 0.006  &  2  & 0  & $7.2 \times 10^{30}$ & 1.1897e+22 & 0.96  & 0.04  &   0   &   0   &   0   \\
 0.006  &  2  & 30 & $7.2 \times 10^{30}$ & 1.1897e+22 & 0.943 & 0.047 & 0.008 & 0.001 &   0   \\
 0.006  &  2  & 45 & $7.2 \times 10^{30}$ & 1.1897e+22 & 0.382 & 0.038 & 0.299 & 0.021 & 0.259 \\
 0.006  &  2  & 60 & $7.2 \times 10^{30}$ & 1.1897e+22 &   0   &   0   & 0.004 & 0.004 & 0.993 \\
 0.012  & 1.1 & 0  & $4.2 \times 10^{30}$ & 2.3833e+22 &   1   &   0   &   0   &   0   &   0   \\
 0.012  & 1.1 & 30 & $4.2 \times 10^{30}$ & 2.3833e+22 & 0.999 &   0   &   0   &   0   &   0   \\
 0.012  & 1.1 & 45 & $4.2 \times 10^{30}$ & 2.3833e+22 & 0.863 & 0.011 & 0.112 & 0.006 & 0.008 \\
 0.012  & 1.1 & 60 & $4.2 \times 10^{30}$ & 2.3833e+22 & 0.149 & 0.016 & 0.166 & 0.044 & 0.625 \\
 0.012  & 1.5 & 0  & $7.9 \times 10^{30}$ & 2.3833e+22 & 0.997 & 0.003 &   0   &   0   &   0   \\
 0.012  & 1.5 & 30 & $7.9 \times 10^{30}$ & 2.3833e+22 & 0.991 & 0.006 & 0.002 &   0   &   0   \\
 0.012  & 1.5 & 45 & $7.9 \times 10^{30}$ & 2.3833e+22 & 0.543 & 0.015 & 0.212 & 0.024 & 0.207 \\
 0.012  & 1.5 & 60 & $7.9 \times 10^{30}$ & 2.3833e+22 & 0.022 & 0.006 & 0.047 & 0.007 & 0.918 \\
 0.012  &  2  & 0  & $1.4 \times 10^{31}$ & 2.3833e+22 & 0.916 & 0.08  & 0.004 &   0   &   0   \\
 0.012  &  2  & 30 & $1.4 \times 10^{31}$ & 2.3833e+22 & 0.92  & 0.038 & 0.039 & 0.002 & 0.001 \\
 0.012  &  2  & 45 & $1.4 \times 10^{31}$ & 2.3833e+22 & 0.293 & 0.008 & 0.173 & 0.023 & 0.504 \\
 0.012  &  2  & 60 & $1.4 \times 10^{31}$ & 2.3833e+22 &   0   &   0   &   0   & 0.001 & 0.999 \\
\hline
\end{tabular}
\begin{tablenotes}\footnotesize
\item[] \textit{Notes:}
\item[1.] $M_{proj}^{iron}$ is the total mass of projectile iron, which is the same at the beginning and end of the simulation.
\item[2.] $X^{Fe}_{interior}$ is the fraction of projectile iron particles in the interior, at the end of the simulation. Results are similarly reported for the fraction of iron particles in the surface, atmosphere, disk, and escaping material.
\end{tablenotes}
\end{threeparttable}
\label{tbl:iron_distribution}
\end{table}

\end{document}